\documentclass[11pt,preprint]{aastex}
\usepackage{natbib}
\begin{document}
\slugcomment{Submitted to The Astrophysical Journal}
\title{The Sensitivity of Convection Zone Depth to Stellar Abundances: An Absolute Stellar Abundance Scale from Asteroseismology}
\shorttitle{Sensitivity of Convection Zone Depth to Stellar Abundances}

\author{Jennifer L. van Saders and Marc H. Pinsonneault}
\affil{Department of Astronomy, The Ohio State University, 140 West 18th Avenue, Columbus, OH, 43210}
\email{vansaders@astronomy.ohio-state.edu,pinsono@astronomy.ohio-state.edu}

\shortauthors{van Saders \& Pinsonneault}
\shorttitle{Sensitivity of Convection Zone Depth to Abundances}
\begin{abstract}

The base of the convection zone is a source of acoustic glitches in the asteroseismic frequency spectra of solar-like oscillators, allowing one to precisely measure the acoustic depth to the feature. We examine the sensitivity of the depth of the convection zone to mass, stellar abundances, and input physics, and in particular, the use of a measurement of the acoustic depth to the CZ as an atmosphere-independent, absolute measure of stellar metallicities. We find that for low mass stars on the main sequence with $0.4 M_{\odot} \le M \le 1.6 M_{\odot}$, the acoustic depth to the base of the convection zone, normalized by the acoustic depth to the center of the star, $\tau_{cz,n}$, is both a strong function of mass, and varies at the 0.5-1\% per  0.1 dex level in [Z/X], and is therefore also a sensitive probe of the composition. We estimate the theoretical uncertainties in the stellar models, and show that combined with reasonable observational uncertainties, we can expect measure the the metallicity to within 0.15 - 0.3 dex for solar-like stars. We discuss the applications of this work to rotational mixing, particularly in the context of the observed mid F star Li dip, and to distguishing between different mixtures of heavy elements. 

\end{abstract}

\section{Introduction}
A profound transition in our understanding of stars is now underway, and one of the major drivers is the detection of pulsations in large samples of sun-like stars.  Much of the immediate interest and effort has focused on using scaling relationships for pulsations to infer global properties, such as mass, radius, and age.   Many new insights into stellar structure will in fact emerge from our new ability to design experiments: for example, using masses outside of binary systems, or ages outside of star clusters.  However, our deepest insights are likely to emerge from diagnostics of internal structure uniquely available from seismic data.   In this paper we focus on one such property: the depth of the surface convection zone (CZ).  We demonstrate that theory predicts strong mass and composition trends in the depth of the CZ with surprisingly small errors.  This raises the prospect of an absolute seismic abundance calibration and rigorous tests of interiors theory.  Furthermore, we may be able to test more challenging issues, such as deep mixing in the mid-F star lithium dip, or the mixture of heavy elements.

A rich spectrum of non-radial pulsations is observed in the Sun.  The study of solar oscillations, or helioseismology, has yielded  insights into solar and stellar structure.  In the case of the Sun we have spatially resolved information and can reconstruct the speed of sound as a function of depth.  For stars we can only observe global modes \footnote{Modes are specified by the three spherical harmonic quantum numbers: $n$, the overtone, $l$, the degree, and $m$, the azimuthal order. For stars other than the Sun, we can generally detect only modes of low degree ($l \lesssim 3$). }, which makes full sound speed reconstructions impractical.   The low $l$ modes have asymptotic frequency spacings predicted by theory, however, which can yield valuable information about the global properties of stars.  Sharp localized changes in structure will also manifest themselves as deviations from these regular spacings; examples include ionization zones and transitions from radiative to convective energy transport.   We briefly review the former before introducing the latter, which are the main focus of the current work.

The average relationships between frequencies of mode pairs $(l,n)-(l,n-1)$  and $(l,n)-(l+2,n-1)$ are frequently referred to as the large and small frequency spacings, respectively.  The former is related to the mean density, while the latter is a measure of the degree of central concentration, and thus helium content, of main sequence stars.  The small frequency spacing is a potent age diagnostic \citep{ulrich1986}.  A significant advantage of these relationships is that useful information can be extracted from the average differences of many mode pairs, increasing the effective signal to noise.  The frequency of maximum power reflects a competition between the spectrum of turbulence generating the sound waves and the acoustic cutoff frequency.  The cutoff frequency is empirically observed to follow regular scaling relationships, and the combination of $\nu_{max}$ and $\Delta \nu$ can be used to solve for the mass and radius \citep{brown1991, mosser2010, belkacem2011, chaplin2011a}.

With better data it is possible to extract entirely new kinds of information.  It was recognized early on that the solar oscillations could be used to measure the strength of the helium ionization zone, and thus by extension the solar surface helium abundance \citep{gough1984}; \citet{delahaye2006} found a literature average of $0.248 \pm 0.004$.  The surface helium is significantly below the initial levels, of order 0.27, required to reproduce the solar luminosity at the solar age.  This difference can be attributed to gravitational settling of helium and heavy elements \citep{aller1960, noerdlinger1977, bahcall1992, bahcall1995} 

The transition from convective to radiative energy transport induces a discontinuity in the temperature gradient, which in turn produces a characteristic response in the frequencies of the modes which cross this boundary.  This phenomenon can be used to infer the depth of the surface convection zone \citep{christensen-dalsgaard1991}.  The agreement between the theoretically predicted depth and the seismic data was a significant triumph for stellar interiors theory; see for example \citet{bahcall1992}. 

In the solar context these seismic tools can be used for precision tests of astrophysics; for example, the convection zone depth can be measured to a remarkable precision, of order $5 \times 10^{-4}$ \citep{basu2004}.  Scalar constraints on convection zone depth and surface helium can be combined to test the absolute solar metal abundance and mixture \citep{delahaye2006}.  Mixing sufficient to explain the solar lithium depletion \citep{weymann1965, pinsonneault1989} can reduce the effects of settling \citep{richard1996,bahcall2001} and has a characteristic signature in the sound speed profile.   Solar data is even of sufficient quality to detect the signature of metal ionization \citep{basu2008}, which can be used to infer the absolute oxygen abundance.   
In stars it is anticipated that the acoustic radius of the convection zone can be measured to of order 2 \% \citep{ballot2004}.  Measurements of the surface helium abundance are more promising in red giants \citep{miglio2010}, but will be technically more difficult in main sequence stars; similar comments apply to the extent of convective cores.  We therefore focus this study on the surface convection zone depth. 
 
To first order the depth of the surface convection zone is determined by the effective temperature, modified by the surface gravity \citep[see][for a discussion]{pinsonneault2001}.  Stellar metallicity plays a second order role, in the sense that lower bulk metallicity implies a shallower surface convection zone at fixed effective temperature.  Our first priority is therefore to quantify these effects, and determine the robustness of the theoretical predictions.  Mixing and element separation processes can also modify the convection zone depth at a detectable level, although it will clearly be more difficult to detect these more subtle shifts in the seismic properties.  We therefore also explore our ability to empirically diagnose these phenomena.

We begin with a discussion of our methods in Section \ref{sec:methods}. We discuss our models results, both in terms of the overall trends in the depth of the convection zone, as well as their magnitude in comparison to theoretical and observational uncertainties in Section \ref{sec:results}. We discuss further potential uses of the depth of the CZ as a diagnostic and conclude in Section \ref{sec:conclusions}.

\section{Calculation of the Acoustic Depth and Theoretical Errorbars} \label{sec:methods}


Our overall approach is similar to that employed in solar model studies.  We define a reference model calculation which includes our current best estimates of the input physics.  We then calibrate a solar model and run a series of models with different masses and compositions; these define our predicted theoretical trends.  We then perform a comprehensive error analysis to estimate the detectability of these signals.  This includes both theoretical errors (for example, nuclear reaction cross-sections or quantum mechanical opacity calculations) and observational errors (for example, the seismic age and mass of the star being tested).    Errors are not easy to estimate in some cases, such as for opacities.  We therefore use the differences between competing calculations as our measure of these theoretical uncertainties.  We also explore some indirect tests of physics not typically included in interiors models, such as rotationally induced mixing.  In a full calculation one would properly account for nonlinear effects rather than employing a strict parameter variation study.   We briefly discuss some such cases, but our main concern is with laying out the baseline theoretical expectations and their overall reliability.  More complex nonlinear calculations would be a logical next step and would be better motivated in the presence of a significant observational database.

\subsection{Calculation of the Acoustic Depth to the Convection Zone} 

We choose to focus on the acoustic depth of the convection zone, $\tau_{cz}$, rather than its physical depth, since $\tau_{cz}$ is the asteroseismic observable of the CZ location. In general, the ``acoustic depth'' is defined \citep{gough1990} as,
\begin{equation}
 \tau=\int_{R_{1}}^{R_{2}}{\displaystyle\frac{dR'}{c_s}},
\end{equation}

where $R$ is the radius and $c_s$ is the sound speed. In order to easily compare $\tau_{cz}$ for stars across a wide range of masses, we use the ``normalized'' acoustic depth, $\tau_{cz,n}$, given by,
\begin{equation}
\displaystyle \tau_{cz,n}= \frac{\tau_{cz}}{\tau_{\star}}=\displaystyle\frac{\int_{R_{cz}}^{R_{\star}}{\frac{dR'}{c_s}}}{\int_{R=0}^{R_{\star}}{\frac{dR'}{c_s}}},
\end{equation}

where the sound speed is simply $  c_s=\sqrt{\frac{\Gamma_{1}P}{\rho}}.$

Both $P$ and $\rho$ as a function of radius can be obtained directly from a converged stellar model. In general, the adiabatic exponent, $\Gamma_1$, is a combination of the quantities 
\begin{equation} \label{eqn:gamma}
 \Gamma_1 = \left(\displaystyle \frac{d\ln{P}}{d\ln{\rho}} \right)_{ad} = \frac{c_p}{c_v}\chi_{\rho},
\end{equation}
where $c_v$ and $c_p$ are the specific heats at constant volume and pressure, respectively, and are related by  
\begin{equation} \label{eqn:cv}
 c_v = \displaystyle c_p - \frac{P}{\rho T}\frac{\chi_{T}^2}{\chi_{\rho}},
\end{equation}
where the derivatives $\chi_{T}$ and $\chi_{\rho}$ are defined as
\begin{equation} \label{eqn:chit}
 \chi_T = \displaystyle{\left(\frac{\partial \ln{P}}{\partial \ln{T}}\right)}_{\rho}, 
\end{equation}
\begin{equation} \label{eqn:chirho}
   \chi_{\rho} = \displaystyle{\left(\frac{\partial \ln{P}}{\partial \ln{\rho}}\right)}_{T}.
\end{equation}
In the case of our stellar models, the values of $\Gamma_1$ come either from tabulated equations of state, or from a combination of $c_p$ and the derivatives $\chi_T$ and $\chi_{\rho}$, which can be calculated within the evolution code itself. 

\subsection{Standard Input}\label{sec:standard_physics}

We define a standard set of input physics from which we create our fiducial stellar models. Models include both helium and heavy element diffusion, since heavy elements sink with respect to lighter elements in a gravitational potential. We use the procedure of \citet{thoul1994}, which numerically solves the full set of the \citet{burgers1969} equations for a multicomponent fluid with no restriction on the number of species considered. For the purposes of computing diffusion coefficients, we treat all heavy elements in the same manner as fully ionized iron \citep[see][]{bahcall1995}. In reality, the effects of diffusion and settling are modified by at least two other physical processes: radiative levitation and mixing. The first is a small effect for the mass ranges we consider, and our models therefore include no prescription for levitation (see Section \ref{sec:discussion} for discussion). We do, however, account for mixing. Mild envelope mixing is needed to explain the Li and Be abundances of low mass stars \citep{pinsonneault1997}. \citet{richard1996} and \citet{bahcall2001} found that mixing sufficient to explain the observed Li depletion has the primary seismic effect of reducing the efficiency of element segregation. We therefore set the diffusion coefficients to 0.8 as in \citet{delahaye2006} to account for the effects of mixing. 

The atmosphere and surface boundary conditions are given by the \citet{kurucz1997} model atmosphere tables \footnote{Models are available at:http://kurucz.harvard.edu/ }. The convection zone depth is only weakly sensitive to the choice of boundary conditions \citep{bahcall1992,delahaye2006}. We choose the Kurucz tables over those of \citet{allard2000} because of their finer gridding in composition. We utilize the recently updated nuclear reaction rates of \citet{adelberger2011} with weak screening \citep{salpeter1954}, and employ the mixing length theory of convection \citep{cox1968, vitense1953}. Opacities are from the Opacity Project (OP) \citep{mendoza2007} for a \citet[][hereafter GS98]{grevesse1998} solar mixture, and are interpolated for each composition as needed. These are supplemented with the low temperature opacities of \citet{ferguson2005}, also for the GS98 mixture. The GS98 solar abundances are in good agreement with asteroseismology \citep[see][]{bahcall2005,basu2008} in comparison to the more recent solar mixture of \citet{asplund2009}, and are thus our default choice. We discuss the effects of adopting the \citet{asplund2009} mixture in Section \ref{subsubsec:zero}. The structural effects of rotation and convective overshoot are neglected in the standard models, although we address such processes further in Sections \ref{subsubsec:sys} and \ref{sec:discussion}. Semiconvection \citep[see][]{kippenhahn} is nominally included, although is of little importance over the stellar mass range we consider. 

We utilize the updated 2006 OPAL equation of state (EOS)\footnote{updated 2006 tables available at http://opalopacity.llnl.gov/opal.html} \citep{rogers1996,rogers2002} and the \citet{saumon1995} EOS for temperature and density combinations outside of the OPAL tables.  To calculate the sound speed throughout each stellar model in a thermodynamically consistent fashion, we use the published values of $\Gamma_{1}$ in the OPAL 2006 EOS. $\Gamma_{1}$ is determined using the values for $P$ and $T$ from the converged model at the end of each timestep, for each shell of the model. Values of $\Gamma_{1}$ for the envelope and atmosphere are likewise acquired directly from the OPAL EOS tables, using the values of $P$ and $T$ from the envelope integration. For our purposes, we neglect the acoustic thickness of the atmosphere in the calculation of $\tau_{cz,n}$, because the modes of interest are generally evanescent in this region (but see Section \ref{sec:caveats} for further discussion of the atmosphere). The base of the convection zone, $R_{cz}$, is defined to be the location where the Schwarzschild criterion for convective instability, $\nabla_{rad}>\nabla_{ad}$, is fullfiled. With the values of $\Gamma_{1}$ obtained from the OPAL 2006 EOS, and the structure from the interior and envelope calculations within the code, we perform the integral $\tau_{cz,n}$ by interpolating the calculated values of $c_{s}(R)$ onto an even grid in $R$, and the integrating the tabulated values using a five-point Newton-Cotes integration formula. 

We use a solar calibration to set the value of the mixing-length parameter, $\alpha$ (the ratio between the convective mixing length and pressure scale height), and the initial composition, $X$, $Y$, and $Z$ such that a $1 M_{\odot}$ model at 4.57 Gyr \citep[see][]{bahcall1995} recovers the solar radius, luminosity, and surface abundance of $R_{\odot}=6.9598\times10^{10} \textrm{ cm}$, $L_{\odot}=3.8418\times10^{33} \textrm{ ergs}\textrm{ s}^{-1}$,  and $Z/X = 0.02289 $ from GS98, respectively. A calibration using this standard set of physics yields $\alpha = 1.93271$, $X = 0.710040$, and $Z =0.018338 $. 

\subsubsection{Composition grid}
We created a larger grid of models for masses 0.4 $M_{\odot}$ - 1.6 $M_{\odot}$ and initial abundances $-1.2 \le \textrm{[Z/X]}\le +0.6$ where $\textrm{[Z/X]} = \log_{10}{\left(\textrm{(Z/X)}_{model} / \textrm{(Z/X)}_{\odot, i}\right)}$ and $\textrm{(Z/X)}_{\odot, i} = 0.025828$, as opposed to the GS98 surface abundance of 0.02289 (the difference being due to element diffusion). We normalize to this initial solar Z/X throughout the paper, which amounts to a zeropoint offset of 0.0524 dex between a metallicity scale normalized to initial versus surface solar abundances. We use models with the standard set of physics to investigate the effect of composition on the location of the convective boundary. The mass range is chosen to roughly coincide with the onset of fully convective models on the low-mass end and vanishingly thin convective envelopes on the high-mass end. The choice of metallicities is motivated by the typical distribution we expect to observe in a sample of field stars. Models are evolved until they leave the main sequence, or until 14 Gyr has elapsed, whichever occurs first. For stars with $M \gtrsim 1.3 M_{\odot}$, the convective envelope becomes less massive than the default fitting point ($1.24 \times 10^{-4} M_{\odot}$) between the interior and envelope solutions. The fitting point is moved to a minimum mass of $1\times10^{-7} M_{\odot}$ to accommodate these models. The grid is composed of models spaced every 0.02 $M_{\odot}$ in mass, 0.2 dex in [Z/X], with initial helium mass fractions of 0.24, 0.26, and 0.28 and $Y_{i,\odot}$ for each combination of mass and metallicity. The result is a grid of some $\sim2400$ models.

\subsection{Theoretical Errorbars on the Acoustic Depth} \label{subsec:theerror}

We examine the theoretical errorbars on the acoustic depth through comparisons of pairs of model grids. The first grid contains ``standard'' models in the sense that they represent the results for the set of input physics described in Section \ref{sec:standard_physics}. Comparison grids are identical to the standard grid except for a single alteration to the input physics. Both grids are subject to separate solar calibrations. We divide the parameter variations into several distinct classes, based on the nature of variation. Some parameters, such as the diffusion coefficient or nuclear cross sections have well defined and random errors. Changes to other parameters, such as the EOS and opacities, can also shift the location of the convection zone in either direction, but the uncertainties in $\tau_{cz,n}$ incurred from switching between different EOS or opacity tables are systematic in nature. We treat the changes induced in $\tau_{cz,n}$ due to well-motivated variations of these parameters as effective $2\sigma$ errorbars on $\tau_{cz,n}$ \citep[as in][]{bahcall1992}. There are also uncertainties that arise simply from our inability to measure the mass, radius, composition, and age of real stars with perfect accuracy, which we will describe as ``observational'' in nature.  A final class, which we will term ``zeropoint uncertainties'' is related to assumptions such as the heavy element mixture or the presence of rotational mixing, and for these cases, the resultant theoretical uncertainties are asymmetric. For example, one can make $\tau_{cz,n}$ smaller by including mixing, but never larger. In these cases, we incur uncertainties in the zeropoint of our relations between various physical parameters and $\tau_{cz}$. These are considered separately from the systematic, random, and observational error sources throughout the paper, but are discussed here for completeness. We proceed, then, to address the uncertainties from each of these sources in turn. The values of $X,Y,Z, \textrm{ and } \alpha$ for each of the physics variations are listed in Table \ref{tbl:errorbar}.

\subsubsection{Random uncertainties}

Element diffusion in stars allows heavier atoms to sink relative to lighter ones. The effect of diffusion is to situate metals, which are a significant source of opacity, deeper within the star than they would otherwise be, resulting in a deeper convective boundary than in models with no element diffusion. The presence of element diffusion in the Sun produces a $1.7\%$ effect \citep{bahcall2001} on the location of the convection zone in Solar models \citep[see also][]{basu2000, bahcall2004}. We construct a calibrated set of models with the helium and heavy metal diffusion coefficients altered by 15\% \citep{thoul1994}, to mimic uncertainty in the strength of diffusion in the interior. Apart from the differing solar calibrations and adjustemnt of the diffusion coefficients, these models are identical to those run with standard physics.

We address the effect of the nuclear reaction rates on the stellar structure, and adopt error estimates for nuclear reaction cross-sections from \citet{adelberger2011}. The major reactions considered are the primary pp chain reactions $\textrm{S}_{1,1}$ (pp), $\textrm{S}_{3,3}$ $(\textrm{He}^{3} + \textrm{He}^{3})$, $\textrm{S}_{3,4}$ $(\textrm{He}^{3} + \textrm{He}^{4})$, and CNO cycle $\textrm{S}_{1,14}$ $(\textrm{P} + \textrm{N}_{14})$, which were each changed by $\pm4\sigma$.

\subsubsection{Systematic uncertainties} \label{subsubsec:sys}

We chose a different prescription for the equation of state in an effort to quantify the change in $\tau_{cz,n}$ due to quantum mechanical uncertainties. We use the \citet{saumon1995} (SCV) EOS instead of the OPAL 2006 EOS \citep{rogers2002} chosen for our standard set of models. We choose this particular variation since the differences between earlier versions of the OPAL EOS are small \citep[see][]{bahcall2004}, and the relative simplicity of the Yale EOS \citep{guenther1992}, which treats the interior as fully ionized and solves the Saha equation for the envelope, is a poor representation of the the state of the art to which EOS calculations have progressed. The OPAL and SCV equations of state are both sufficiently modern, and yet have very different approaches to the problem, and so the SCV EOS serves as a useful comparison. Because we draw the value of $\Gamma_1$ in the standard models directly from the EOS tables, we must alter the manner in which we calculate $\Gamma_1$ for the SCV EOS models. Using the relationships between the derivatives $\chi_{\rho}$, $\chi_{T}$ and the specific heat at constant pressure, $c_p$ (Equations \ref{eqn:gamma} ,\ref{eqn:cv}, \ref{eqn:chit}, \ref{eqn:chirho}), calculated numerically within YREC, we combine these values to calculate $\Gamma_1$ and the sound speed throughout the star.

There are two primary sources for high temperature opacities in stellar interiors models, the OPAL group \citep{rogers1996} and the Opacity Project (OP) \citep{badnell2005}, each of which approaches the quantum mechanical calculation of the high-temperature opacities in a fundamentally different way. A thourough comparison and discussion of the differences between the two methods is present in \citet{seaton2004}. The differences in the Rosseland mean opacity for the conditions found at the base of the solar convection zone are of order 5\% \citep{seaton2004}. The opacity plays a significant role in determining the precise location of the base of the convection zone \citep{bahcall2001}, and so we test the sensitivity of $\tau_{cz,n}$ to our choice of opacity table by running variant models using the OPAL opacities, instead of our default choice of the OP opacities.  

We expect that the choice of atmosphere and boundary conditions will be most important for very cool stars. We test the worst-case dependence of $\tau_{cz,n}$ on the choice of atmosphere boundary condition by creating a calibrated grid of models for a grey atmosphere boundary condition. We note that this exercise only quantifies the dependence of $\tau_{cz,n}$ on the boundary condition, since the portion of $\tau_{cz}$ due to the atmosphere, $\tau_{atm}$, is neglected in the calculation of $\tau_{cz,n}$.

We also consider the importance of convective core overshoot \citep[see][]{zahn1991, maeder1975} to the determination of $\tau_{cz,n}$. While in principle overshoot in all convective layers is possible, and affects the local composition, we consider convective core overshoot in particular, because the added fuel supplied to the core through overshoot-related mixing could have broader impacts on the physical structure of the star. The addition of core overshoot should primarily affect the high end of our mass range, where models begin to develop convective cores. We choose a core overshoot parametrized by the pressure scale height, with a value of 0.2 pressure scale heights. We do not enforce an adiabatic gradient in the overshoot zone or add envelope 'undershooting'; overshooting is treated purely as 'overmixing', and the thermal structure is left unchanged. Observationally, were significant overshooting to be present, observers would see it as an effective change in the location of the convective boundary. Depending on the nature of the overshooting, this could manifest itself as either a zeropoint or mass-dependent shift in the location of the CZ. Here we consider only the manner in which overshoot induced mixing affects the evolution because of increased fuel supply to the core.

\subsubsection{Observational Uncertainties} \label{subsubsec:obs}

We can expect uncertainties in stellar parameters such as $T_{eff}$, $M$, $R$, $\bar{\rho}$, $\tau_{cz,n}$, $Y$, and the age simply from the nature of the observations with which they are determined. We likewise consider our assumption of a particular mass-radius relationship as a form of observational uncertainty. In all cases, these are uncertainties dictated by our ability to \emph{measure} stellar properties, and are therefore of a fundamentally different nature than the uncertainties described above. We also note that these uncertainties are subject to potential rapid improvement, typically on a time scale shorter than those for improvements in opacity tables or equations of state. 

Asteroseismic age diagnostics are sensitive to the helium fraction in the stellar core, and thus provide information about how far along its main sequence lifetime a star has progressed \citep{ulrich1986}. \citet{creevey2009} suggests that we can obtain the ages of main sequence stars to within 10\% of the MS lifetime, and already \citet{metcalfe2010} present an asteroseimic age for KIC 11026764 accurate to 15\% with currently existing Kepler data. Future missions, such as GAIA, which aim to attain precise parallaxes on a large sample of stars, may eventually allow us to better constrain the age based on an absolute luminosity, but we proceed with the assumption that age can be measured to this 10\% accuracy, and propogate these uncertainites through our models.

The abundance of helium in stars is notoriously difficult to measure directly and represents another source of observational uncertainty. Since luminosity is also a function of the helium content, constraints on the mass and the luminosity (at fixed $X$,$Z$, and age) lead to constraints on the helium, an idea that goes as far back as \citet{schwarzschild1946}. If we take $L = 4\pi R^2 \sigma T_{eff}^4$ and assume reasonable measurement uncertainties in $R$ and $T_{eff}$, the uncertainty in $L$ is $\sigma_{L}^2 =\displaystyle \left( \frac{2 \sigma_{R}}{R} \right)^2 + \left( \frac{4 \sigma_{T_{eff}}}{T_{eff}} \right)^2$, and the uncertainty in $Y$ due to that in $L$ is $\sigma_{Y}^2 = \sigma_{L}^2 \displaystyle \left( \frac{\partial Y }{\partial L}\right)^2$. Finally, the uncertainty propagated to $\tau_{cz,n}$ is then $\sigma_{\tau}^2 = \sigma_{Y}^2 \displaystyle \left( \frac{\partial \tau}{\partial Y}\right)^2$. We calculate these derivatives numerically from our composition grid. We assume that $R$ can be measured to a fractional uncertainty of 2\% and $T_{eff}$ to within 100K.

A portion of the observational uncertainties will come directly from the asteroseismic measurements themselves, namely our ability to precisely constrain the large frequency separation and glitch signatures. We assume that the value of $\bar{\rho}$ is attainable to within a relative uncertainty of 1\% from asteroseismic measurements \citep{verner2011}, and that $\tau_{cz,n}$ can be measured to within 2\% \citep{ballot2004}.

We have assumed for our standard grid that a single mixing-length parameter, calibrated for a solar model, is valid over the entire range of masses and compositions we consider. Because we employ only the mixing length theory of convection in these models, we have no ability to test how $\tau_{cz,n}$ changes as a result of different theoretical assumptions about convection, and rather we choose to view variations in $\alpha$ as an uncertainty in the mass-radius relationship (but see Section \ref{sec:discussion} for a further discussion of convection theory). Because the general approach in stellar modeling is to determine the mass and calibrate the model such that the correct radius is recovered, we treat this as an observational error. To test how severely this may impact the inferred depth of the CZ, we construct a grid of many different values of $\alpha$, and choose $\alpha (M)$ such that for all masses considered, the radius of the star in the ``altered physics'' grid is about 1\% larger than in the single, solar calibrated $\alpha$ case in the standard grid. Observed discrepancies from the theoretical mass-radius relationship are observed to be as large as 10\% for cool, low-mass stars in binary systems where they can be well studied \citep{kraus2011, irwin2009, bayless2006}. On both the high and low mass ends of the mass range, the stellar radii become rather insensitive to changes in the mixing-length parameter. For high mass stars, this is because the pressure scale height is small enough that large changes in $\alpha$ itself are physically of little significance. On the low mass end, because stars are nearly fully convective, changes in $\alpha$ tend to shift stars along the main sequence, rather than changing the relation. In both extremes, no change in $\alpha$ is ever sufficient to produce a 10\% difference in the radius. Our chosen $\Delta{R} = 0.01$ is achievable over nearly the entire mass range considered with sufficient changes to the value of $\alpha$. Although scaling these theoretical errorbars to larger uncertainties in the stellar mass-radius relation is not unreasonable, it is important to note that for stars with $M \lesssim 0.6 M_{\odot}$ and $M \gtrsim 1.4 M_{\odot}$, the model radius becomes insenstive to $\alpha$ and simple scalings will fail.

\subsubsection{Zeropoint uncertainties} \label{subsubsec:zero}

Observational and theoretical evidence suggests that some form of mixing operates in both the Sun and other stars, and that this mixing can have effects on the apparent efficiency of diffusion. We know from modelling of the Sun that diffusion alone does not adequately reproduce solar light-element depletion relative to meteorites \citep{richard1996, bahcall2001}, and that rotationally induced mixing provides a well-motivated physical process by which the observed depletion could be achieved \citep{pinsonneault1989}.  \citet{balachandran1995} finds that diffusion alone cannot explain the Li abundances in M67, and it is generally believed to be the signature of some form of deep mixing. \citet{deliyannis1998} likewise finds correlated Li and Be depletion patterns in Hyades F-stars \citep[the strong Li depletion first recognized by][]{boesgaard1986} is best supported by a slow mixing of stellar material. The primary consequence of mixing for our purposes is the accompanying decrease in the efficiency of element diffusion, possibly to the point to which it appears that diffusion does not operate at all. Recent observations of NGC 6397 \citep{korn2006} and theoretical modelling efforts \citep{chaboyer1992,dotter2008} found that diffusion must be partially suppressed in order to explain the observed trends in the light elements of metal-poor stars. To mimic the effects of strong mixing, we construct models with no helium or heavy element diffusion.

The mixture of heavy elements, even from the Sun, is another important systematic error source. We consider here both the case of revised Solar heavy element abundances, and the case of $\alpha$-element enhancement in low-metallicity models. In both cases, the relative abundances of important contributors to the opacity are altered with respect to iron, and we seek to investigate the sensitivity of $\tau_{cz,n}$ to changes in the element mixtures..  

In the case of the Sun, there exists a well known tension between the solar CZ depth and sound speed profile inferred from helioseismology versus that implied by the recent solar abundances of \citet{asplund2004,asplund2009}. These recent abundances are based on non-LTE, 3D radiative transfer calculations of the solar atmosphere and represent the state-of-the-art in modern atmosphere modeling. However, solar models constructed with this new, low bulk metallicity are in worse agreement with seismic diagnostics such as the surface helium abundance, CZ depth, and solar sound speed profile than models with the older GS98 mixture \citep{bahcall2005, basu2008}.  Although the new solar mixture has a similar iron abundance as the old GS98 mixture, the CNO elements are significantly adjusted, and the oxygen abundance in particular is quite low. Because these elements tend to be completely ionized in the deep interior of stars, they contribute little in the way of opacity in the core, but have significant opacities near the location of the base of the CZ in solar-like stars \citep{delahaye2006}. It is important to note that similar work on the Solar mixture by \citet{caffau2011}, also using a 3D analysis, arrived at a higher oxygen and bulk metallicity than \citet{asplund2009}. Although future work may alleviate the conflict between helioseismic inversions and the \citet{asplund2009} mixture, we choose to investigate how the lowest published oxygen abundances affect our conclusions regarding the depth of the convection zone. We construct a calibrated set of models for the \citet{asplund2009} mixture, using OP \citep{seaton2005} and \citet{ferguson2005} low temperature opacity tables adjusted for the change in mixture. The models are calibrated to a surface abundance of $\textrm{Z/X} =0.0199 $. We note that while the pre-main sequence stellar models and opacity tables are adjusted to reflect the difference in abundance pattern, the EOS and atmosphere tables are not. In both cases the correct mass fraction in metals is used, and the errors incurred from the difference in mixture in the atmosphere and EOS should be negligible, since the change of mixture primarily affects the CNO elements and therefore nuclear burning and the highly metal-sensitive opacities. 

In the case of metal-poor halo stars, we may also expect that there may be deviations from the Solar mixture due to different chemical enrichment histories. 
We consider $\alpha$-enhanced models, with [$\alpha$/Fe] = +0.2 \citep[following][]{dotter2008}. We compare models with the standard GS98 mixture at [Z/X], [Fe/H] = $-1.0$ to $\alpha$-enhanced models with [Fe/H] = $-1.0$ but [Z/X] = $-0.85$ with $Y_i = 0.247$ in both cases. As with the Solar case, both low and high temperature opacity tables and initial models are adjusted for the change in mixture. We supply EOS and atmospheres with the correct bulk metal abundances, but relative abundances are not adjusted.

\subsubsection{Combined Uncertainties}
We calculate the magnitude of the uncertainties from each of the error sources above for an assumed reference model at solar composition, with an age of 5 Gyr, with the standard set of physics. In combining the errors from each source, we treat the uncertainties as uncorrelated.  We add individual sources of error in quadrature. For example, the random error on $\tau_{cz,n}$, $\sigma_{\tau,rand}$ is composed of $\sigma_{\tau, rand}^2 = \Sigma \sigma_{nuc}^2  + \sigma_{diff}^2$, where the individual errors are from the nuclear reaction rates and diffusion coefficients, respectively. Similar combinations of error terms are calculated for each class of uncertainty, random, systematic, and observational. The total uncertainty on $\tau_{cz,n}$ is
\begin{equation}
 \sigma_{\tau}^2 = \sigma_{\tau, rand}^2 + \sigma_{\tau, sys}^2 + \sigma_{\tau, obs}^2,
\label{eqn:sigtau}
\end{equation}
 where the random, systematic, and observational errorbars are added in quadrature. We consider the zeropoint uncertainties separately, since they are of a fundamentally different nature, and have asymmetric effects on the depth of the convection zone.

Although an exhaustive investigation of the cross terms in our theoretical error estimates are beyond the scope of this paper, we do comment briefly on a few cases in which we have investigated the role of crossterms with the age uncertainties. The shapes of the curves in $\tau_{cz,n} - T_{eff}$ and $\tau_{cz,n}-\bar{\rho}$ space depend strongly on age, and so we have singled out this particular error source in which to look for crossterms. We find that for the case in which both the age and helium are simultaneously considered, that the error on $\tau_{cz,n}$ from the combination of age and Y uncertainties is negligible when the interplay between the error sources is considered. Crossterms between age uncertainties and changes in the physics are likewise negligible. The exception is in the case of overshoot models, in which the time dependence of the diffusion and the age uncertainties interact, inflating the errorbars by up to 30\% for the high mass models when both age and overshoot uncertainties are considered simultaneously. This is not unreasonable, since overshooting is both more important for more massive objects, and affects the amount of fuel in the stellar core.  In general, we recommend that one estimate errors, and potential cross-terms, on a source-by-source basis when actual data are available.

\section{The sensitivity of the convection zone to mass and composition} \label{sec:results}

In this section we present the main predictions of our models, the results of which are given in Table \ref{tbl:model_grid}. The depth of the convection zone is a strong function of mass and composition, as predicted by interiors theory and recovered in our models. With our theoretical errorbars, we find that these trends are a strong, observable effect, even when we account for uncertainties in both the interiors models and observationally derived quantities. The mass and composition dependencies can be viewed in terms of the purely asteroseismic quantities of the mean density ($\bar{\rho}$, here given as $\bar{\rho} = M/R^3$) and normalized acoustic depth $\tau_{cz}/\tau_{\star} = \tau_{cz,n}$, and any analysis can in principle rely on solely asteroseismically obtained measurements. Furthermore, the composition dependence of $\tau_{cz,n}$ is more pronounced in $\bar{\rho}$ - $\tau_{cz,n}$ space, as opposed to, for example, $T_{eff}$ - $\tau_{cz,n}$ space: the use of purely asteroseismic variables is not only useful, but beneficial. Finally, we show that with appropriate observations, various tests of stellar physics, beyond the basic question of the location and presence of a convection zone, are possible.

\subsection{Mass dependence of $\tau_{cz,n}$} \label{subsec:mass}
\begin{figure}
	\centerline{\includegraphics[scale = 0.9]{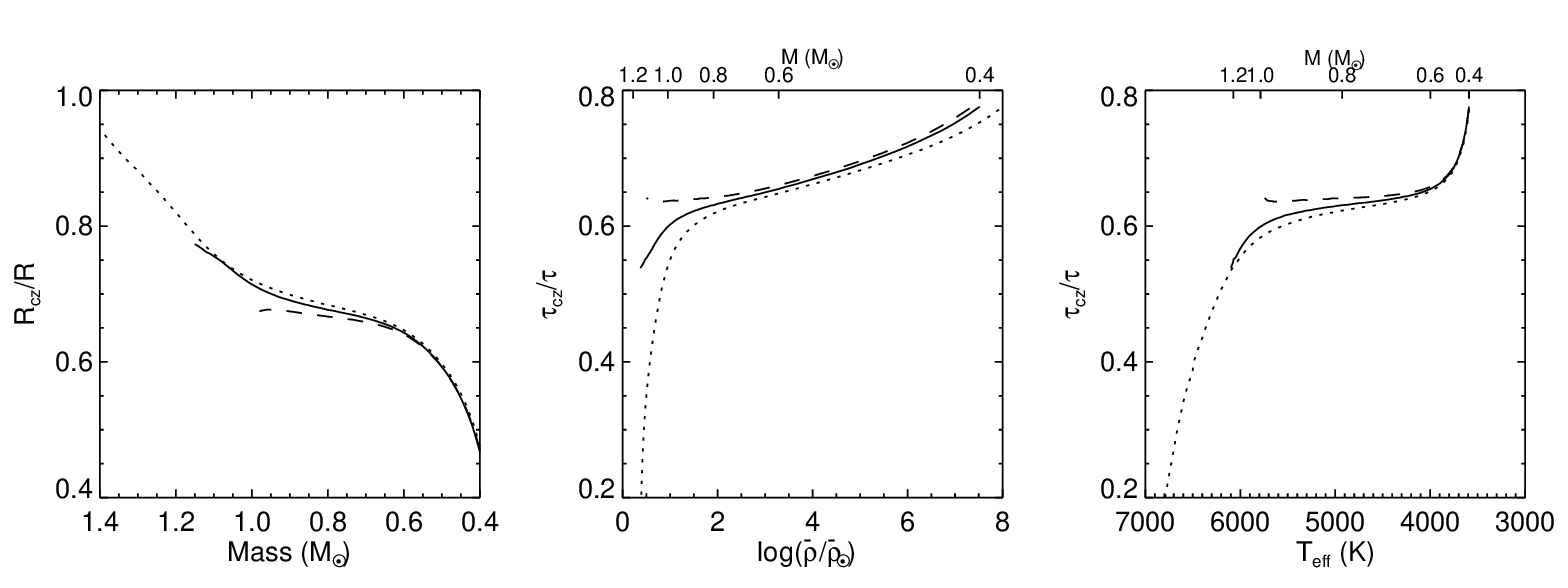}}
       \caption{Left panel: The physical depth of the convection zone, normalized by the radius of the star, as a function of mass. Middle panel: The acoustic depth to the convection zone, normalized by the acoustic depth from surface to center of the star. Right panel: The normalized acoustic depth as a function of effective temperature. The dotted curve is for solar composition models at 1.0 Gyr, the solid for 5.0 Gyr, and dashed for 10.0 Gyr. All models are on the main sequence, with the central hydrogen fraction $X_c \ge 0.0002$. The top axis in the center and right panels gives the mass for objects at 5.0 Gyr (solid curve). The corresponding discussion can be found in Section \ref{subsec:mass}.}
	\label{fig:czmapping}
\end{figure}

We expect from very simple interiors arguments that the depth of the convection zone must be a strong function of mass. To first order the location of the base of the convection zone is set by the location of the H and He ionization zones, where the adiabatic temperature gradient is suppressed below the radiative temperature gradient and the criterion for convection is satisfied. This is illustrated in Figure \ref{fig:czmapping}, which shows $R_{cz}/R_{\star}$ versus mass on the left, and $\tau_{cz,n}$ versus $\bar{\rho}$ in the middle and $\tau_{cz,n}$ versus $T_{eff}$ on the left for standard models at solar composition and an age of 1,5 and 10 Gyr. The expected strong mass dependence of the depth of the CZ is clearly present, and the mapping from the $R_{cz}/R_{\star}$ vs. $M$ to  $\tau_{cz,n}$ vs. $\bar{\rho}$ planes is a simple one, which preserves the sense of the trends. Models with relatively high masses have vanishingly thin, shallow convection zones, whereas low mass models are nearly fully convective. If our physical understanding of what sets the location of the convection zones is correct, observations of many different stars should display this strong predicted trend \citep[also discussed in][]{monteiro2000}. In the case in which this strong dependence is not observed, we immediately learn that there is some fundamental physical process that has been neglected or poorly treated in interiors models. 

We can also comment here on the basic time dependence of such relationships. At very young ages, the entire mass range we consider ($0.4 \le M_{\odot} \le 1.4$) is on the main sequence. The basic shape of the $\tau_{cz,n}$ vs. $\bar{\rho}$ changes very little between 0.5 and 1.0 Gyr, for example, because the most massive objects we consider have main sequence lifetimes of a least 1 Gyr. Once we begin to look at later times, however, the low density tail of the curves begins to show significant changes, because the stars that occupy that part of parameter space are progressively less massive (more massive models have evolved off the main sequence and out of our realm of consideration). The mean density of a star decreases over the course of its main sequence lifetime, so it is possible for an older, less massive star to have the same mean density as a young, more massive star. Their convection zones, however, generally not at same relative depth on the main sequence, and so the low density tail of a given $\tau_{cz,n}$ vs. $\bar{\rho}$ curve shifts to deeper convection zones at late times. We incorporate this feature of the relation into our observational error estimates in later sections.

\subsection{Composition dependence of $\tau_{cz,n}$} \label{subsec:comp}

\begin{figure}
	\centerline{\includegraphics[scale = 1.0]{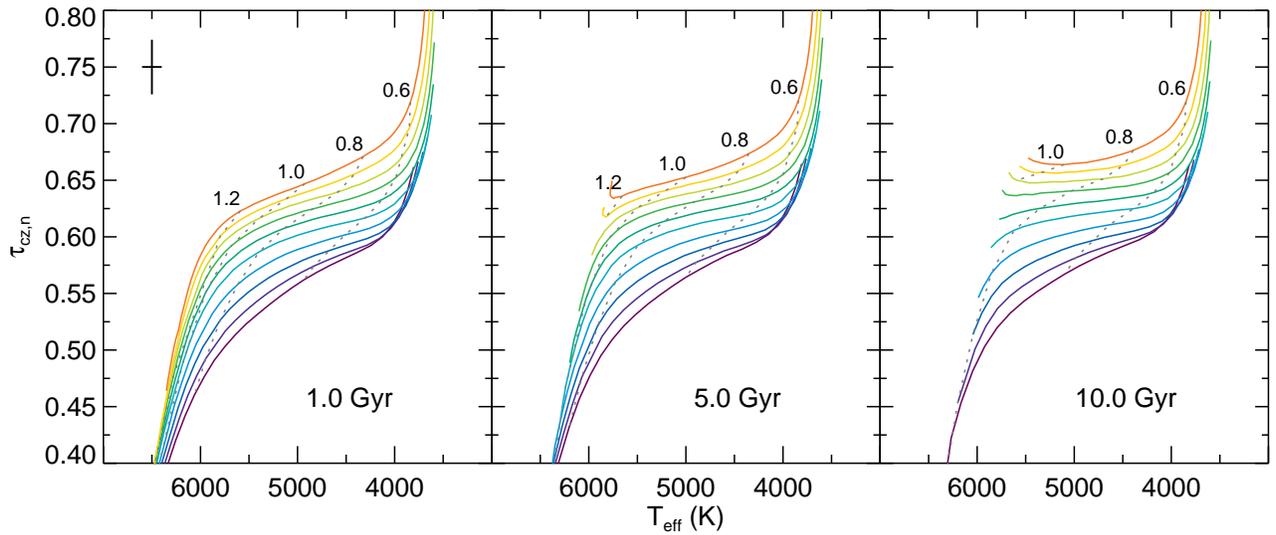}}
       \caption{The variation of the normalized acoustic depth of the base of the convection zone as a function of composition and $T_{eff}$. Solid lines represent models of fixed initial [Z/X] (reference to the \emph{initial} solar abundance) from red/top most (+0.6) to purple/bottom most ($-1.2$) spaced every 0.2 dex. The composition dependence of the acoustic depth is plotted for three representative ages, and dotted lines are plotted in gray for constant mass, in solar units. All models are stars on the main sequence, with $X_{core} \ge 0.0002$ and have an initial, solar-calibrated helium of $Y_i = 0.271$. Representative observational error bars on the quantities $\tau_{cz,n}$ and $T_{eff}$ are shown in the left most panel. See Section \ref{subsec:comp} for discussion.}
	\label{fig:standard_teff_1}
\end{figure}

\begin{figure}
	\centerline{\includegraphics[scale = 1.0]{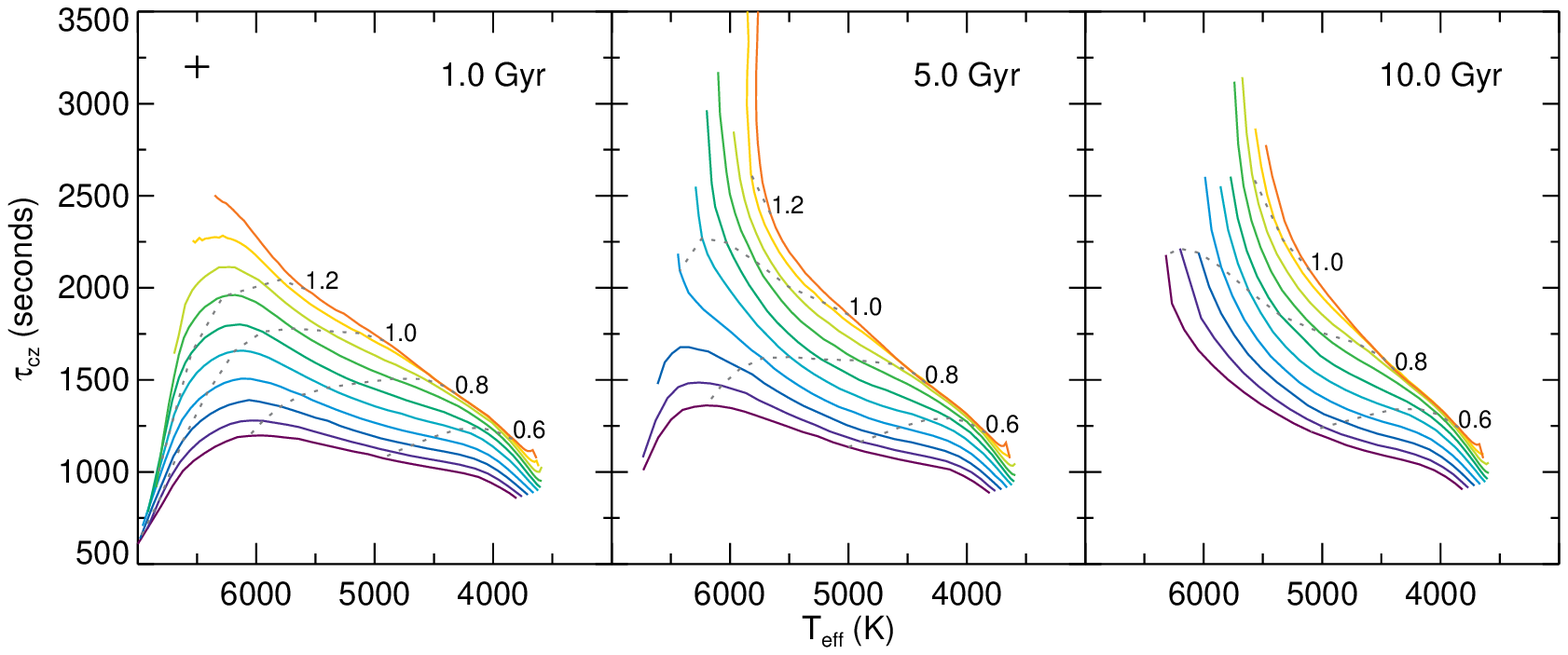}}
       \caption{The variation of the absolute acoustic depth (in seconds) of base of the convection zone as a function of composition and $T_{eff}$. Solid lines represent models of fixed initial [Z/X] from red/top most (+0.6) to purple/bottom most ($-1.2$) spaced every 0.2 dex. The composition dependence of the acoustic depth is plotted for three representative ages, and dotted lines are plotted in gray for constant mass, in solar units. All models are stars on the main sequence, with $X_{core} \ge 0.0002$ and have an initial, solar-calibrated helium of $Y_i = 0.271$. Representative observational error bars on the quantities $\tau_{cz}$ and $T_{eff}$ are shown in the left most panel. See Section \ref{subsec:comp} for discussion. }
	\label{fig:standard_teff_2}
\end{figure}

\begin{figure}
	\centerline{\includegraphics[scale = 1.0]{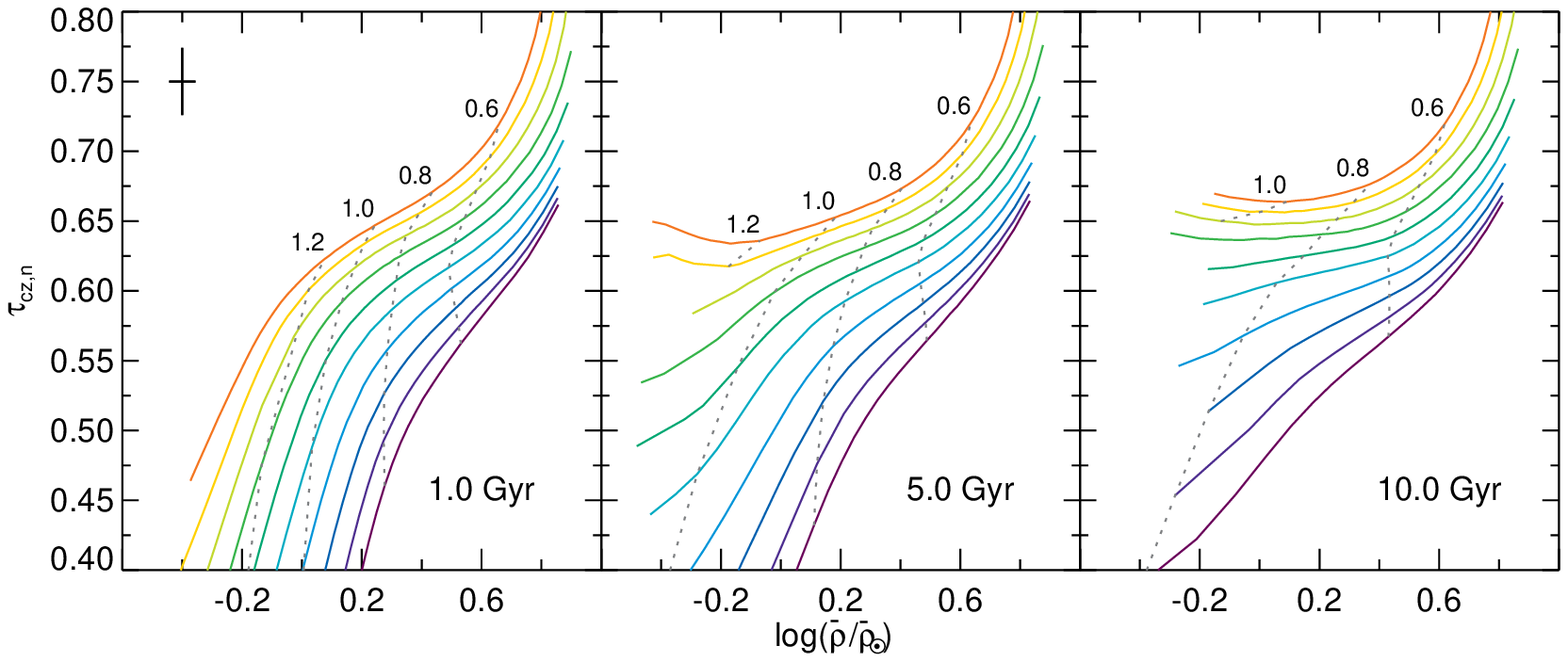}}
       \caption{The variation of the normalized acoustic depth of the base of the convection zone as a function of composition and $\bar{\rho}$. Solid lines represent models of fixed initial [Z/X] from red/top most (+0.6) to purple/bottom most ($-1.2$) spaced every 0.2 dex. The composition dependence of the acoustic depth is plotted for three representative ages, and dotted lines are plotted in gray for constant mass, in solar units. All models are stars on the main sequence, with $X_{core} \ge 0.0002$ and have an initial, solar-calibrated helium of $Y_i = 0.271$. Representative observational error bars on the quantities $\tau_{cz,n}$ and $\bar{\rho}$ are shown in the left most panel. See Section \ref{subsec:comp} for discussion. }
	\label{fig:standard_rho_1}
\end{figure}

\begin{figure}
	\centerline{\includegraphics[scale = 1.0]{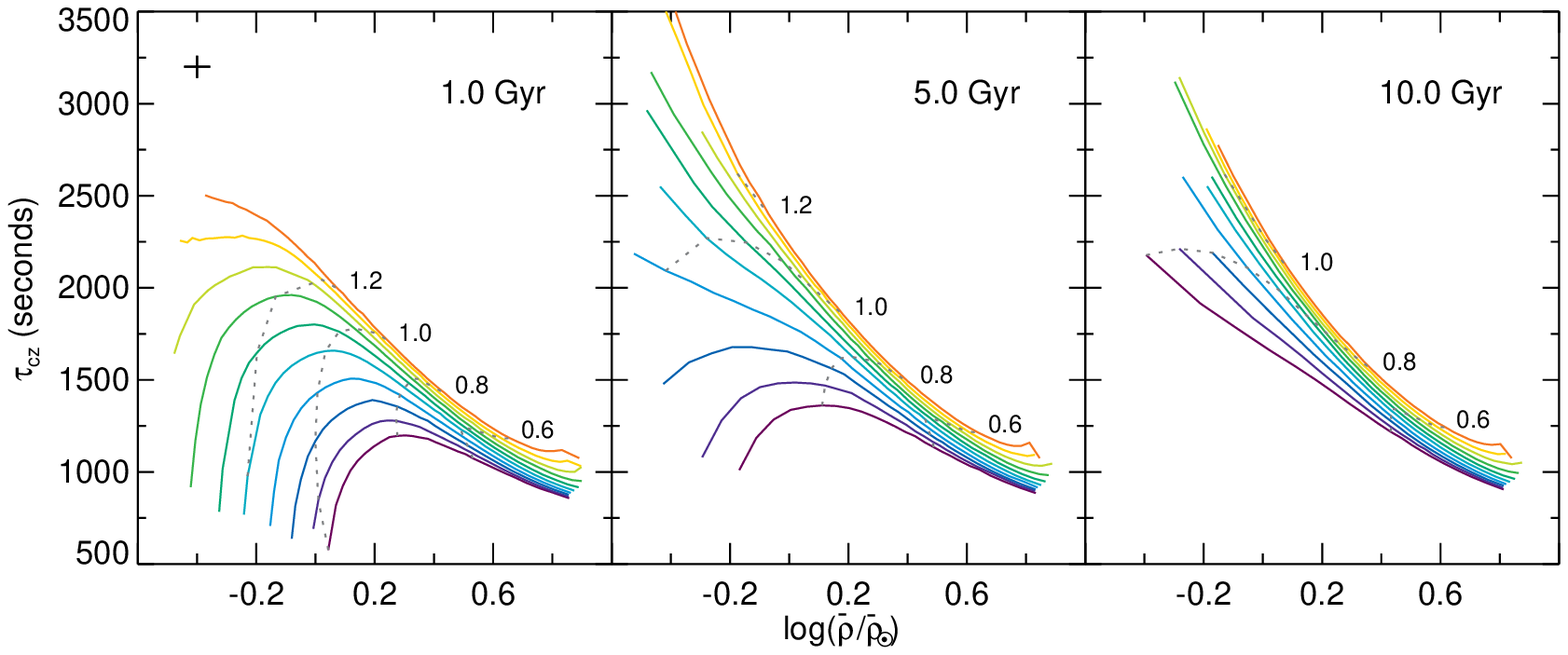}}
       \caption{The variation of the absolute acoustic depth (in seconds) of base of the convection zone as a function of composition and $\bar{\rho}$. Solid lines represent models of fixed initial [Z/X] from red/top most (+0.6) to purple/bottom most ($-1.2$) spaced every 0.2 dex.  The composition dependence of the acoustic depth is plotted for three representative ages, and dotted lines are plotted in gray for constant mass, in solar units. All models are stars on the main sequence, with $X_{core} \ge 0.0002$ and have an initial, solar-calibrated helium of $Y_i = 0.271$. Representative observational error bars on the quantities $\tau_{cz}$ and $\bar{\rho}$ are shown in the left most panel. See Section \ref{subsec:comp} for discussion. }
	\label{fig:standard_rho_2}
\end{figure}

While mass should be the primary determinant of the depth of the convection zone, it is clear from similarly simple arguments that the composition should also play some role in the location of the convection zone. The radiative temperature gradient is given by \citet{kippenhahn} as $\nabla_{rad} =  \frac{3}{16\pi a c G} \frac{\kappa l P}{m T^4}$ 
where $a,c$ and $G$ are the usual physical constants, $l$ is the luminosity of the shell with mass $m$, $P$ and $T$ the pressure and temperature, and $\kappa$ the opacity. With all else being equal, an increase in the opacity leads to an increase in the radiative temperature gradient, which in turn means that the criterion for convection is satisfied deeper (at higher T) within the star. While metals are an almost negligible fraction of the mass, because they contribute significantly to the opacity and are important in determining the precise location of the convection zone. This is indeed what we find, shown in Figures \ref{fig:standard_teff_1},\ref{fig:standard_teff_2}, \ref{fig:standard_rho_1}, and \ref{fig:standard_rho_2}, where $\tau_{cz,n}$ and $\tau_{cz}$ (in seconds) is plotted with respect to $\bar{\rho}$ and $T_{eff}$. We predict that $\tau_{cz,n}$ changes by 0.5-1\% per 0.1 dex in [Z/X] over most of the mass range we consider. Furthermore, this composition signature is an \emph{absolute}, rather than relative, measure of stellar abundances. It is most pronounced in the $\tau_{cz}$ - $\bar{\rho}$ plane, rather than in the $\tau_{cz}$ - $T_{eff}$ plane. Therefore, the most preferable space in which to work is also the space in which the composition measurement can be made solely with asteroseismically obtainable variables: spectroscopic and photometric characterization of the stellar parameters is only necessary as an additional constraint on the stellar parameters.

\begin{figure}
	\centerline{\includegraphics[scale = 0.9]{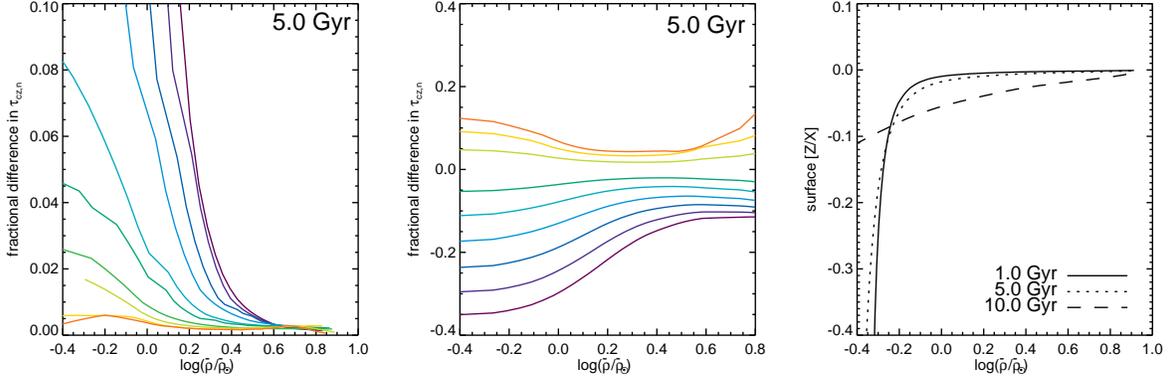}}
       \caption{Left panel: The fractional difference in $\tau_{cz,n}$ between the surface abundance and initial composition at \emph{constant} [Z/X]. This difference arises because of gravitational settling of heavy elements, and would manifest itself as a $T_{eff}$ dependence of the surface [Z/H] in in mono-composition sample. Center panel: Lines of iso-composition in surface abundance, compared to a reference model at solar surface abundance. Left panel: surface abundance as a function of $\bar{\rho}$ for models of solar composition. The solid line denotes models at 1 Gyr, the dotted at 5 Gyr, and the dashed at 10 Gyr. See Section \ref{subsec:comp} for discussion. }
	\label{fig:surfaceZH}
\end{figure}

Because our models include gravitational settling, heavy elements tend to sink relative to light ones, and the surface [Z/X] is generally not the same as the initial abundance, and changes as a function of time. Figure \ref{fig:surfaceZH} shows both the difference between the initial and surface abundances. This difference arises because of gravitational settling of heavy elements and would manifest itself as a $T_{eff}$ dependence of the surface [Z/H] in a sample with homogeneous initial composition, such as an open star cluster.  If we consider models at fixed $\bar{\rho}/\bar{\rho_{\odot}} = -0.2$ at 1.0, 5.0, and 10.0 Gyr as in the right panel of Figure \ref{fig:surfaceZH}, we find that the difference between the surface and initial abundance is most pronounced for the 10.0 Gyr curve. This is a balance between two competing factors: at earlier times, more massive stars with short settling timescales are still on the main sequence and occupy this density range. At later times, less massive stars occupy this region and have longer settling timescales, but longer MS lifetimes over which settling can occur. This difference between the initial and surface abundances is important for any comparison of asteroseismic and atmospheric abundance measurements: the value of the surface abundance for a given model is a physics and age dependent property.

\begin{figure}
	\centerline{\includegraphics[scale = 0.9]{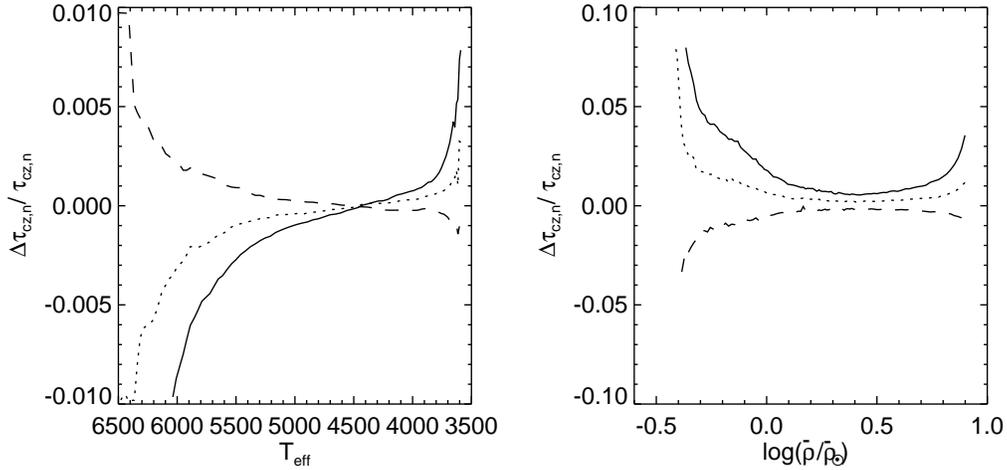}}
       \caption{The fractional difference in $\tau_{cz,n}$ between models with an initial solar helium abundance ($Y=0.271$) and $Y=0.24$ (solid), $Y=0.26$ (dotted) and $Y=0.28$ (dashed) at constant initial Z/X (solar) for models of 1.0 Gyr. $\Delta \tau_{cz,n} / \tau_{cz,n} > 0$ represents models in which the convection zone is deeper than in the standard, solar $Y$ case. See Section \ref{subsec:comp} for discussion.}
	\label{fig:helium}
\end{figure}

We also show in Figure \ref{fig:helium} the fractional difference in $\tau_{cz,n}$ among models of different initial helium abundances at constant, solar Z/X at 1.0 Gyr. We factor the uncertainty in the helium, as discussed in Section \ref{subsubsec:obs} into our observational error budget.

\subsection{Uncertainties in the relationship of $\tau_{cz,n}(\bar{\rho}, T_{eff})$} \label{subsec:uncertainties}

\begin{figure}
	\centerline{\includegraphics[scale = 1.1]{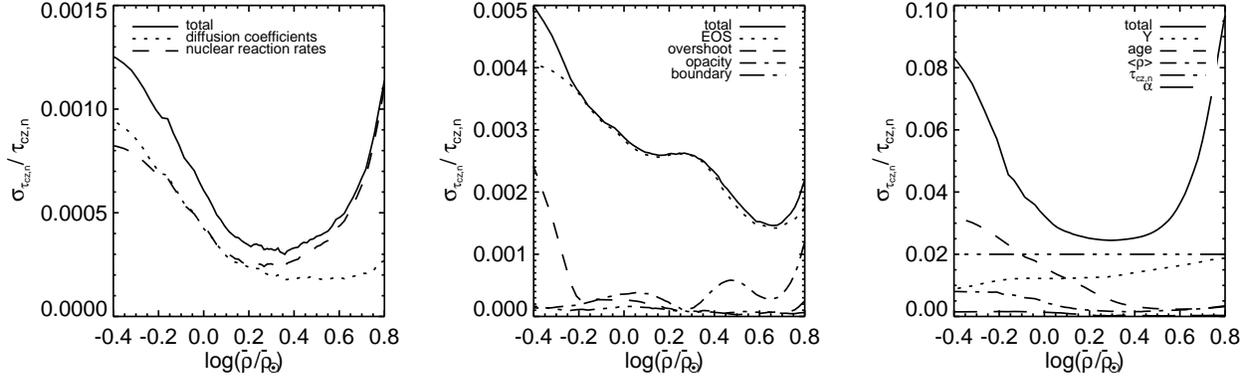}}
       \caption{The fractional uncertainty in the measurement of $\tau_{cz,n}$ for a 5 Gyr old, solar composition star from our each class of uncertainties, from left to right: random, systematic, and observational. See Section \ref{subsec:theerror} for a discussion of the uncertainty calculations. The sources of uncertainty are coded as follows: left panel: solid line: total, dotted: diffusion coefficients, dashed: nuclear reaction rates. Center panel: solid- total, dotted- EOS, dashed-overshoot, dot dashed-opacity, double-dot dashed-boundary conditions. Right panel: solid-total, dotted-Y, short dashed- age, dot dashed-measurement of $\bar{\rho}$, triple-dot dashed- measurement of $\tau_{cz,n}$, long dashed-mass-radius relation, through changes in $\alpha$.     }
	\label{fig:error}
\end{figure}

The characterization of the uncertainties in the relationships among mass, composition, and $\tau_{cz,n}$ means that we can not only comment on the existence of an important trend, but also quantify whether it is presently observable. We first discuss the results of the variant models introduced in Section \ref{subsec:theerror}, and then show that these uncertainties are small enough that the depth of the CZ can be used as a precise indicator of composition. We present representative uncertainties in $\tau_{cz,n}$ due to random, systematic, and observational uncertainties in Figure \ref{fig:error} for models with a solar composition and age of 5 Gyr. 

The contribution to $\sigma_{\tau}$ from sources such as the diffusion coefficients and nuclear reaction rates is below 0.1\%, except on the very low and very high mass ends of the distribution. Contributions from the systematic class of errors are somewhat more significant, with the EOS being the most important source of uncertainty in this particular grouping. As expected, the uncertainties due to overshoot and the choice of boundary condition are small, below 0.05\%. 

Uncertainties incurred from errors in our knowledge of the global properties of the star ($M, T_{eff}, R, Y$ and age) are by far the largest source of error in $\tau_{cz,n}$. At low mean densities, uncertainty in $\tau_{cz,n}$ due to uncertainty in the age is the most significant contributor to the observational error, which is unsurprising given the behavior of the curves in Figure \ref{fig:czmapping} as a function of age. At small masses the uncertainty in the age itself is large because the objects have very long MS lifetimes and we can only measure the age asteroseismically to within 10\% of that lifetime. However, the change in $\tau_{cz,n}$ with time is comparatively small and so the uncertainties in $\tau_{cz,n}$ induced by age variations are modest for the low-mass models. For higher masses, our ability to measure the age asteroseismically is substantially better, but the shape of the the $\tau_{cz,n}$ vs. $\bar{\rho}$ is changing significantly with time because models are evolving off the main sequence and a given density probes very different masses at different times. Therefore, the age-induced uncertainties are largest on the high mass (low density, high $T_{eff}$) parts of the curve. 

One should note that the uncertainties due to the mass-radius relationship are relatively small, but that we have also chosen a very modest $\Delta R = 1\%$. In principle, these errors can be scaled for larger radius uncertainties in the mid-mass range. On both the large and small mass extremes, however, simple scalings of radius errors using the mixing length $\alpha$ will fail (as mentioned in \ref{subsubsec:obs}). On the low mass end, even a very large change in $\alpha$ has only a small effect on $\tau_{cz,n}$. For high masses, however, large changes in $\alpha$ also lead to substantial changes in $\tau_{cz,n}$. Therefore, one must be wary if attempting to scale these particular errorbars for larger radius discrepancies for the higher mass stars.

The uncertainties due to unknown helium abundances are also non-negligible contributors to the observational error. On average, given the assumed uncertainties in our ability to measure luminosity, we can hope to constrain the helium mass fraction to within 0.01-0.02, using the technique described in Section \ref{subsubsec:obs}. Additional information from parallaxes or asteroseismic determinations of the helium could better constrain this number. It is however, encouraging that we will be able to assign realistic errorbars to the helium, as opposed to \emph{ad hoc} estimates.

\begin{figure}
	\centerline{\includegraphics[scale = 0.9]{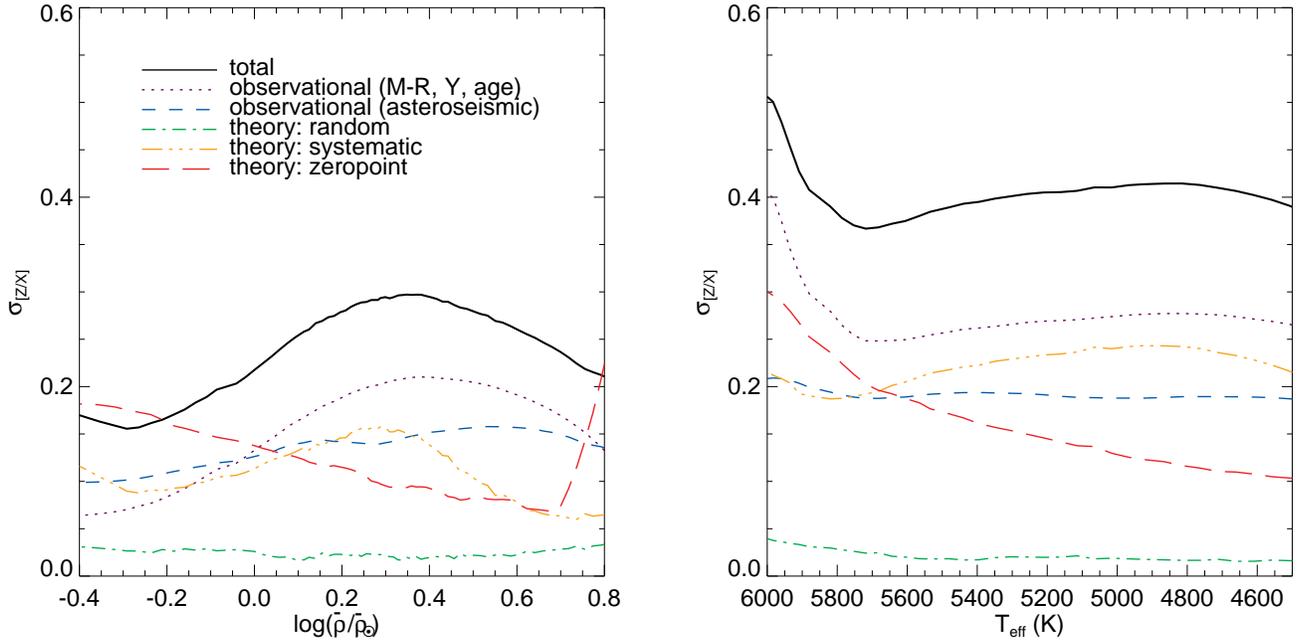}}
       \caption{The uncertainty in the measurement of the metallicity of a solar composition, 5 Gyr old star using $\tau_{cz,n}$ as a composition indicator. The dashed (blue) line represents the contribution from asteroseismic uncertainties on the observables, the dashed (purple) from observational uncertainties on the mass-radius relation, Y, and age, the triple-dot-dashed (orange) from systematic uncertainties, long-dashed (red) the zeropoint uncertainties (mixture + mixing), dot-dashed (green) the random uncertainties, and solid (black) the combined systematic, random and observational uncertainties. See Section \ref{subsec:uncertainties} for discussion.}
	\label{fig:sigzh}
\end{figure}

We combine this suite of systematic, random and observational uncertainties on $\tau_{cz,n}$ and translate this uncertainty into a measure of our ability to measure [Z/X] through $\sigma_{\textrm{[Z/X]}}^2 = \sigma_{\tau_{cz,n}}^2 \displaystyle \left(\frac{\partial [Z/X]}{\partial \tau_{cz,n}}\right)^2$, where the components of $\sigma_{\tau_{cz,n}}$ are shown in Equation \ref{eqn:sigtau}. Figure \ref{fig:sigzh} shows $\sigma_{\textrm{[Z/X]}}$ for each error source, systematic, random, and observational, as well as the combined total error. The result is that $\tau_{cz,n}$ is actually strikingly sensitive to composition, even when we include reasonable theoretical and observational errors. The uncertainty in [Z/X] is in the range of 0.15-0.3 dex over the mass range we consider.

In addition, we alter the compositions and perform the same comparisons for sets of metal-poor and metal-rich models (rather than errors derived for the solar case presented in Figure \ref{fig:sigzh}). We assume a simple chemical evolution scheme of the form
\begin{equation}
 Y = Y_p + \frac{dY}{dZ}Z
\end{equation}
with $Y_p = 0.246$ and $dY/dZ = 1.0$. We take models with \emph{initial} Z/X ratios at a tenth solar and 2.5 times solar, with the change in $Y$ determined by our chemical evolution assumptions. Examining the fractional differences in $\tau_{cz,n}$ for sets of models at different initial abundances allow us to check whether our ability to determine [Z/X] depends strongly on the composition.  We find that the situation is quite favorable, with $\sigma_{\textrm{[Z/X]}} \approx 0.2-0.3$ for the brightest $\textrm{[Z/X]}_i = -1.0$ models at 10 Gyr (representative of a halo star population) when all assumptions about the observational errors are identical to those in the solar example. In this case of a metal rich object with $\textrm{[Z/X]}_i = +0.4$, $\sigma_{\textrm{[Z/X]}}$ is similar to that in the solar case. This suggests that $\tau_{cz,n}$ remains a good indicator of composition across the entire regime of compositions we have considered, provided our assumed observational errors remain representative.

\subsection{Probing the physics of stellar interiors with $\tau_{cz,n}$ diagnostics}

Our analysis suggests several interesting tests of the conditions that prevail in stellar interiors using measurements of $\tau_{cz,n}$, beyond the potential to constrain composition and confirm basic theoretical predictions of interiors models. Because the depth of the convection zone has some sensitivity to the particular physical assumptions of interiors models, we can invert the question confronted above and ask: if we can measure $\tau_{cz,n}$, and if we can trust our stellar parameters derived by means other than asteroseismology (photospheric metallicities in particular), can we use the patterns we observe in $\tau_{cz,n}$ to infer something about the physics of the interior?  In this section we first describe how one could use measurements of $\tau_{cz,n}$ as a test of the physics responsible for the observed mid-F star Li depletion. Secondly, we outline the manner in which one could use a large sample of $\tau_{cz,n}$ measurements to constrain the stellar abundance pattern.

\subsubsection{The Li dip} \label{subsubsec:lidip}

We observe that Li undergoes a severe depletion event in stars of roughly $6200-6350K$ \citep{boesgaard1986, balachandran1995}. While the existence of the Li dip is well established observationally, theory has yet to converge upon a mechanism responsible for the effect. Many different mechanisms have been proposed to explain Li depletion in both the Sun and other stars: through mixing by waves, mass loss, diffusion, and rotationally induced mixing \citep[see][and references therein, for a thorough discussion]{pinsonneault1997}. We focus here on rotational mixing and imagine the following scenario: if stars undergo an episode of strong rotational mixing at about $\sim6350K$, and that mixing has the effect of completely erasing the element segregation induced by gravitational settling and diffusion, then we expect a jump in $\tau_{cz,n}$ as the model crosses the Li dip boundary and the underlying physical assumptions change. We find that the no-diffusion models have values of $\tau_{cz,n}$ that are $\sim6\%$ lower than those in the standard model over the temperature range of 6200-6350 K for models at 1 Gyr (at later times, few models populate this temperature range). We imagine a scenario in which stars are well represented by the standard model curve up until the edge of the Li dip, at which point they undergo a strong mixing event which erases the effects of diffusion, and the star abruptly jumps onto a no-diffusion model curve. If we consider the case in which we observe pairs of stars, one of the Li ``peak'' at 6200 K, the other in the Li ``dip'' at 6350 K, the important quantity to consider is the slope, $\Delta \tau_{cz,n}/ \Delta T_{eff}$ of the standard models over this tempertaure range, compared to that between a standard model on the low temperature side of the Li dip, and a no-diffusion model at the high temperature side. We consider the slope  $\Delta \tau_{cz,n}/ \Delta T_{eff}$ of the standard model and the scatter in that slope present when we introduce our aforementioned changes to the physics, compared to the slope between the standard and no-diffusion models over the dip. In the ideal case in which we have perfect measurements of $T_{eff}$ and $\tau_{cz,n}$, and the only uncertainties are theoretical (not observational), then the jump in the value of $\tau_{cz,n}$ across the Li dip is an $8\sigma$ event. However, when observational errorbars due to age, $Y$, and astereoseismic measurement uncertainties are included on the standard-to-no-diffusion model slope, the jump in $\tau_{cz,n}$ is significant at the $0.8\sigma$ level per pair of stars In $\tau_{cz,n}-\bar{\rho}$ space the significance is slightly decreased, due to the fact that the mapping between $T_{eff}, \bar{\rho}$ and mass changes slightly between the standard and no diffusion cases, and conspires in this plane to make the jump less visible. We conclude then, that with a sample of $\sim 15$ pairs of stars, if a mixing event is responsible both for removing the signatures of diffusion and providing the means to deplete Li, then a trend in the observed values of $\tau_{cz,n}$ should be visible at the $3\sigma$ level.

There clearly exist some caveats to this prediction, the most important of which is that it is unclear whether rapidly rotating stars of the sort on the hot side of the Li dip will actually display solar-like oscillations and produce reliable measurements of $\tau_{cz,n}$. The detection and interpretation of solar oscillations in rapidly rotating stars is still among the principal challenges in asteroseismology \citep{reese2010}. Furthermore, our models have negelected the structural effects of rotation, and we have chosen to use no-diffusion models as a simple representation of rotationally mixed stars. Rotationally induced changes to the structure, especially in hot, rapidly rotating Li dip stars, may be important. Nevertheless, we provide a useful test of the mixing in stellar interiors in the temperature range of the Li dip.

\subsubsection{ Variant Elemental Mixtures}

We briefly also consider $\alpha$-enhanced models, with [$\alpha$/Fe] = +0.2 \citep[following][]{dotter2008} at fixed [Fe/H] at 10 Gyr, to mimic an old halo star population. If one were to consider a pair of stars at the same $\bar{\rho}$, and assume that the theoretical uncertainties on $\alpha$-enhanced stellar models are the same as in the case of a solar mixture, then their measured values of $\tau_{cz,n}$ should be different by of order 1.0-1.3$\sigma$ for $\bar{\rho}<\bar{\rho_{\odot}}$ with the inclusion of observational uncertainties. In general, the difference between models with solar and $\alpha$-enhanced mixtures is $\Delta \tau_{cz,n} \sim 0.03-0.06$ for the stars with $\bar{\rho}< \bar{\rho_{\odot}}$ at 10 Gyr, which will also be the most likely objects to be detected in missions such as \emph{Kepler}. Therefore, if one can measure the [Fe/H] values for several pair of stars at the same age and mean densities, then it may be possible to distinguish between solar and $\alpha$-enhanced mixtures on the basis of the normalized acoustic depth to the CZ. This is another example of the power of pairwise comparison, since the theoretical errors effectively cancel for two stars of the same mean density and age, and it is only observational errors that effect the significance of the difference in $\tau_{cz,n}$.

As we discussed in Section \ref{subsubsec:zero}, the recent \citet{asplund2009} oxygen abundances are contentious in part because the revision implies a solar CZ depth that does not agree with asteroseismic measurements. We investigate here whether we can utilize ensemble measurements of $\tau_{cz,n}$ to learn about the oxygen abundance relative to the total metal abundance of other stars. This question is well-posed in an open cluster situation, in which the stars are of uniform age and composition, and the stellar parameters are somewhat better constrained than in the case of a random field star. The typical difference between standard solar models and Asplund mixture models is $\Delta \tau_{cz,n} \sim 0.005-0.01$ for models at solar composition at 1 Gyr. 
We will focus on a sample of stars, randomly drawn from a uniform distribution in $5500 \le T_{eff} \le 6500$, each with an average combined observational and theoretical uncertainty of $\sigma_{\tau} \sim 0.03$ (including a 0.3 dex [Z/X] uncertainty). The quantity $\tau_{standard} - \tau_{observed}$, where $\tau_{standard}$ is the acoustic depth that would be measured for standard physics, and $\tau_{observed}$ is the value measured for stars with an Asplund mixture, quantifies the zeropoint offset induced by the difference in mixture. Given our standard assumptions about the theoretical and observational uncertainties, careful measurements of $\tau_{cz,n}$ for a 25 star sample could detect a mixture difference at $3\sigma$. One should note that this is an idealized example: we've assumed that all stars are exactly the same age and composition, with exactly the same oxygen abundance, and exactly the same physical processes operating within them. In reality, we could imagine that some effect, such as mixing, might operate differently in stars of different masses, which would dilute the abundance pattern signal, or make it appear anomalously strong, depending on the sense of the mass dependence. Both mixing and low relative oxygen abundances tend to make the CZ more shallow, and so disentangling the zeropoint offset due to a different relative oxygen abundance may in practice be quite challenging. This analysis also relies on a correct theoretical zeropoint calibration: our standard physics models must be anchored correctly, because any theoretical zeropoint offset could be mistaken as a real signal. It is currently unclear how well we can achieve this, as even numerical sources of error become important at the $\Delta \tau_{cz,n} = 0.005$ level.  Nevertheless, this is another potentially useful application of careful measurements of $\tau_{cz,n}$. In the solar case, detailed information about the mixture could be obtained through a simultaneous measurement of the surface helium and $R_{cz}$. We anticipate that a similar approach, if practical, will be required in the stellar context.

\subsection{Caveats} \label{sec:caveats}

We discuss the caveats to our findings in regards to our treatment of the atmosphere and convection theory, and other physically important elements of stellar interiors, such as rotation, magnetic fields, and  radiative levitation.

We have neglected the contribution to the acoustic depth from the atmosphere throughout our discussion: we have only considered the acoustic depth due to the interior and envelope portions of our models. Although all models are run with a Kurucz atmosphere boundary condition ($P$ at $T = T_{eff}$), we perform the calculation of the ``acoustic thickness'' of the atmosphere using a grey atmosphere, to allow us to calculate the necessary integrals as a function of radius. While we have already demonstrated that the choice of boundary condition produces small ($\sim 0.5\%$) changes in the normalized acoustic depth, we expect that there are somewhat larger uncertainties associated with the atmosphere itself.  We integrate the sound speed in the atmosphere using the assumption that the change in radius is given by  $dr =\displaystyle \frac{d \tau}{ -\kappa \rho}$ \citep{cox1968}, where $\tau$ is the optical depth, and $\kappa$ the opacity. For models with a grey atmosphere boundary condition, we find that the acoustic thickness of the atmosphere, $\tau_{atm}$ is $50 \gtrsim \tau_{atm} \gtrsim 250$ seconds, with $\tau_{atm}$ increasing with increasing stellar mass. $\tau_{atm}$ is typically 5-7\% of $\tau_{cz}$ for all but the most massive stars with the thinnest convective envelopes, where it is a more significant fraction of $\tau_{cz}$. $\tau_{atm}$ is $\sim 4\%$ of the total acoustic travel time in the interior + envelope regions for all masses. The inclusion of $\tau_{atm}$ in the normalized acoustic depth can change $\tau_{cz,n}$ by up to 20\% for massive objects with thin convective envelopes, but is typically 5\% for stars with $M_{\star} \lesssim 1.0 M_{\odot}$  A calibrated, standard physics, solar model at 4.57 Gyr produces a $\tau_{cz} = 2100 $s, whereas the solar value for $\bar{\tau_{cz}}$, which includes surface and atmospheric contributions is $\sim2200-2300$s \citep{verner2004}, which suggests our model results are in good agreement with reality. From these arguments, we can reasonably expect that the neglect of the atmosphere may result in a few hundred second offset between our models and reality. 

One can also expect that given the nature of our asteroseismic observables, the derived values should not suffer substantially from uncertain surface term corrections. In both the case of the mean density (derived from the large frequency separation) and the acoustic depth to the CZ (derived from an oscillatory signal in what would otherwise be uniformly spaced frequencies) it is only the relative difference in the surface term among the modes that will bias the measurements. Furthermore, in models where surface convection responsible for the asteroseismic surface terms is treated more carefully than in our MLT approach \citep{stein2000}, the inclusion of the additional physics significantly improves agreement with high frequency solar modes, but the solar base of the CZ remained essentially unchanged.

We note, however, that we have assumed an unquantified systematic uncertainty in our choice of a single prescription for solar convection. While the near surface convection appears not to be of great importance to our analysis, we have included no test here of the importance of variant convective theories to our results. 

As is the case with stellar modeling in general, the relative importance of element separation and mixing is one of our primary uncertainties. Throughout our analysis, we have ignored the effects of rotation, except in the decreased efficiency of diffusion encapsulated in the diffusion coefficients. In terms of determinations of $\tau_{cz}$, we expect that the most important contributions to be in the form of adjustments to the relative efficiencies of mixing and diffusion, which we have shown have an impact of $\tau_{cz}$. For rapidly rotating objects, the rotational splitting of the modes and introduction of new modes of oscillation \citep{reese2010} may make mode identification and interpretation challenging. 

Magnetic fields may produces changes in the sound speed near the surface of the star, but we expect only a small correction to the sound speed in the deep interior for all but extremely strong internal magnetic fields. Again, since the modes of importance have turning points well below the photosphere, we expect corrections from magnetic fields to be small. Furthermore, \citet{chaplin2011b} finds that highly active stars are less likely to display detectable solar-like oscillations, which suggests that the primary role of magnetic fields may be in dictating whether we can detect p-modes at all, rather than affecting the acoustic glitch signature itself. 

We have also neglected radiative levitation \citep[see][for discussion]{pinsonneault1997}, which can selectively levitate some elements relative to others. While our analysis captures the impact of global metal diffusion, it does not account for selective levitation of individual elements. In particular, this can affect elements such as iron, which contributes substantially to the opacity. The effects of radiative levitation are most pronounced in hotter stars with thin surface convection zones. The accuracy of the most massive of the models we consider may therefore suffer from our neglect of radiative levitation. 

In general, we advocate pairwise comparisons of measured values of $\tau_{cz,n}$ for stars which one suspects differ significantly in only one way, i.e., testing the mass-$\tau_{cz}$ relation using two stars of very different mass but similar composition and age, or two stars with similar ages and masses but different compositions. Obtaining an accurate zeropoint calibration of this relation is currently challenging. For example, variations in $\tau_{cz,n}$ on the order of 0.005 can be induced due to numerical differences between models with different envelope fitting points. Even in solar models, similar numerical uncertainties due to interpolation can affect the inferred base of the convection zone \citep{bahcall2004}. Physical effects, such as the presence of envelope undershooting, could also appear as a zeropoint offset in the relation. Furthermore, as shown in \citep{bahcall2004}, the interpolation of quantities such as radiative opacity tables is uncertain on the 1-3\% level near the base of the CZ in the Sun. The best approach is therefore to compare pairs of interesting stars, in which case zeropoint calibrations will be of less importance.    

\section{Discussion and Conclusion} \label{sec:discussion} \label{sec:conclusions}

We have discussed the factors that affect the location of the base of the convection zone thoroughly, but have neglected the second (and probably more commonly discussed) source of acoustic glitches in asteroseismic spectra: the helium ionization zone. In the Sun, measurements of the helium ionization zone glitch have constrained the surface helium abundance \citep{basu2004}, and hopes are high that this will also prove possible in the stellar case. We expect that even if the He ionization zone is sensitive to metallicity, the dynamic range of the effect will be much too small to make precise metal abundance measurements. Furthermore, the ionization zone lies in the outermost layers of the star, and is subject to uncertain surface term corrections, much more so than the deeper base of the convection zone. For these reasons a similar analysis on the effects of mass and metal content on the helium ionization zone is beyond the scope of this paper, but we emphasize that the sensitivity is unknown, and it could yet prove to be an interesting diagnostic. 

Our entire analysis has focused only on main sequence stars, which are inherently fainter, and their mode amplitudes are smaller than the subgaints, which have recently proven to be a rich source of asteroseismic information \citep{brand2011, metcalfe2010, bedding2006}, and are among the most common stars with detectable solar-like oscillations in the current Kepler ensemble \citep{chaplin2011a}. A preliminary analysis of models evolved onto the subgiant branch shows similar trends with composition, with composition effects of similar magnitude. At fixed age, $\tau_{cz,n}$ now increases with decreasing $\bar{\rho}$ and probes models increasingly closer to beginning the ascent up the giant branch with deepening convective envelopes. A full analysis of the sensitivity of $\tau_{cz,n}$ to composition in subgiants is underway. If the sensitivity and theoretical errorbars are similar to those on the main sequence, we stand to benefit substantially from extending the analysis to subgiants, which have larger mode amplitudes and higher luminosities, which can help to reduce observational errors. 

This unique means of measuring the composition promises a host of interesting applications. We could, for example, test the tendency of planets to be found around hosts of spectroscopically high metallicity \citep{fischer2005}. Differences between the interiors and atmospheres based compositions could help to constrain whether planets are more likely to be found around intrinsically metal rich stars, or whether planets themselves tend to enrich the outer layers of their hosts with heavy elements. A simple comparison of spectroscopically and asteroseismically determined compositions would in itself be an interesting consistency check, and potentially offer insights into the reliability of both methods and physical processes such as element diffusion. The striking sensitivity of the location of the convection zone to composition even at very low metallicities also provides an interesting and relatively rare insight into the interiors of metal-poor stars. These are only a handful of the numerous ways in which we can begin to use asteroseismic measurements such as this as novel diagnostics of stellar interiors and stellar populations.

To conclude, we have created a grid of stellar models of different compositions and examined the sensitivity of the acoustic depth to the convection zone as a function of composition, mass, and our assumptions about the input stellar physics. We make three primary predictions based on the analysis of our models:
\begin{enumerate}
\item{We predict strong trends in the depth of the convection zone as a function of mass and composition. The asteroseismic CZ depth indicator $\tau_{cz,n}$ can be different by as much as factor of $\sim2$ between stars of masses 0.4 and 1.4 $M_{\odot}$. Composition produces changes in $\tau_{cz,n}$ of order 1\% per 0.1 dex in [Z/X]. $\tau_{cz,n}$ remains sensitive to the composition even at low ($\sim-1.0$) values of [Z/X]. These strong scalings provide a simple test of interiors theory, and an absolute abundance measure independent of atmospheric modelling. Furthermore, the problem is well posed in $\tau_{cz,n}-\bar{\rho}$ space, both of which are purely asterosesimic observables. }

\item{Reasonable estimates of theoretical and observational uncertainties suggest that not only is $\tau_{cz,n}$ sensitive to the composition, but that the uncertainties in the relationship are small. On average, we expect to be able to measure \emph{absolute} abundances to 0.15- 0.3 dex for solar-like stars at 5 Gyr given the assumed observational and theoretical uncertainties.} 

\item{Finally, the measurement of the depth to the convection zone has potential diagnostic power as a means of probing theoretical uncertainties. In particular, we have addressed the manner in which one would use $\tau_{cz,n}$ to test for rotational mixing in Li dip stars, and to test for differences in the relative element abundances in an ensemble of targets.}

\end{enumerate}

Measurements of $\tau_{cz,n}$ have the potential to both constrain interiors theory in terms of the balance between diffusion and mixing, element abundance patterns, and the basic prediction of a strongly mass dependent CZ depth. The technique also and offers a unique, absolute abundance measure, which is inherently useful in the study of the chemical enrichment of the galaxy, and benchmark for comparison to stellar atmosphere derived abundances. This is a powerful tool that can help us to precisely measure stellar parameters and test the physics of stellar interiors.

\section{Acknowledgments}
We would like to thank Travis Metcalfe for helpful discussions throughout this work, and  Franck Delahaye for both discussions and his aid with obtaining the necessary opacity tables. We also thank Jennifer A. Johnson, Sarbani Basu, and Benjamin Shappee for helpful comments along the way. This material is based upon work supported by the National Science Foundation Graduate Research Fellowship under Grant RF\#743796 (JVS), and NASA grant NNX11AE04G (MHP).

\bibliographystyle{apj}

\begin{figure}
	\centerline{\includegraphics[scale = 1.0]{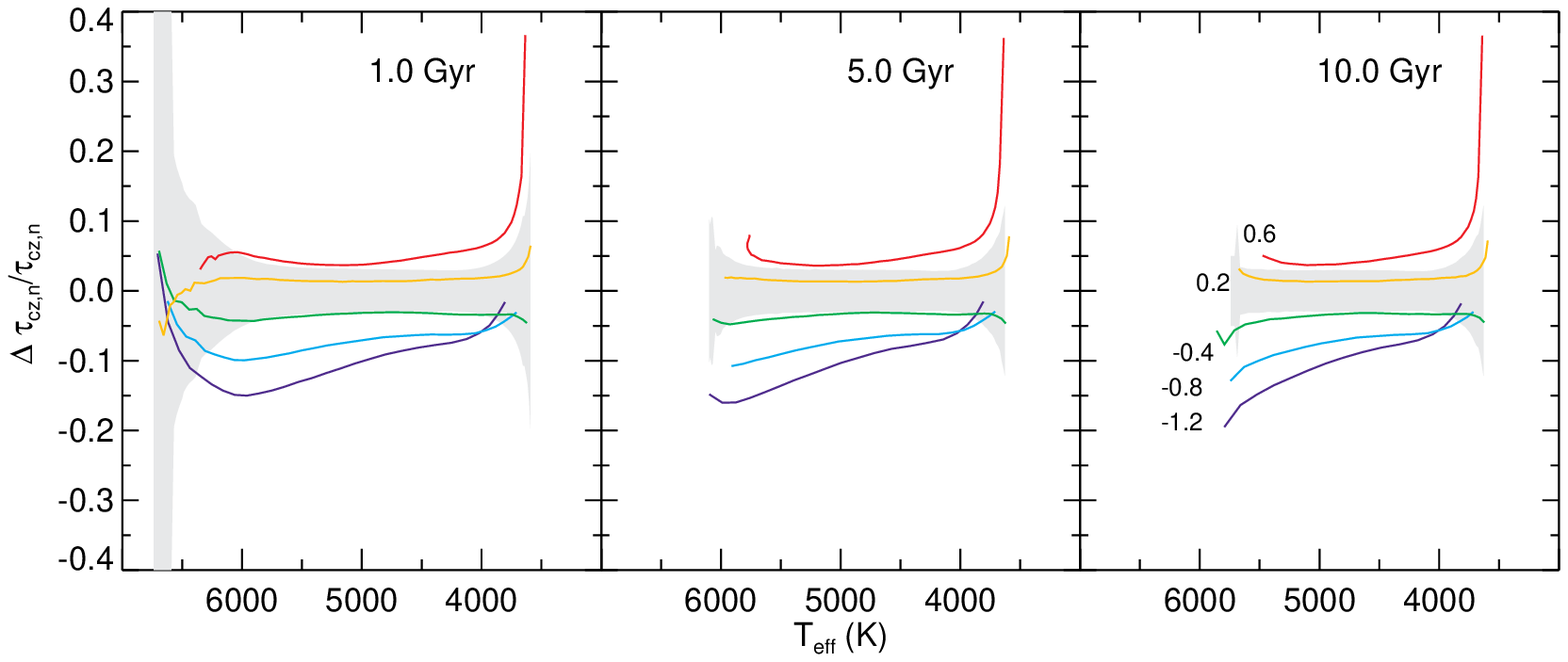}}
       \caption{The fractional difference in $\tau_{cz,n}$ (the normalized acoustic depth to the base of the convection zone) between models of different compositions as a function of $T_{eff}$. Each solid line represents the fractional difference in $\tau_{cz,n}$ between a given [Z/X] ($-1.2$, $-0.8$, $-0.2$, 0.2, 0.6, marked for reference in the right panel) and $\tau_{cz,n}$ for a solar composition model. The gray shaded region represents observational and theoretical errors on $\tau_{cz,n}$, both described in detail in Sections \ref{subsec:theerror} }
	\label{fig:standard_difference_teff}
\end{figure}

\begin{figure}
	\centerline{\includegraphics[scale = 1.0]{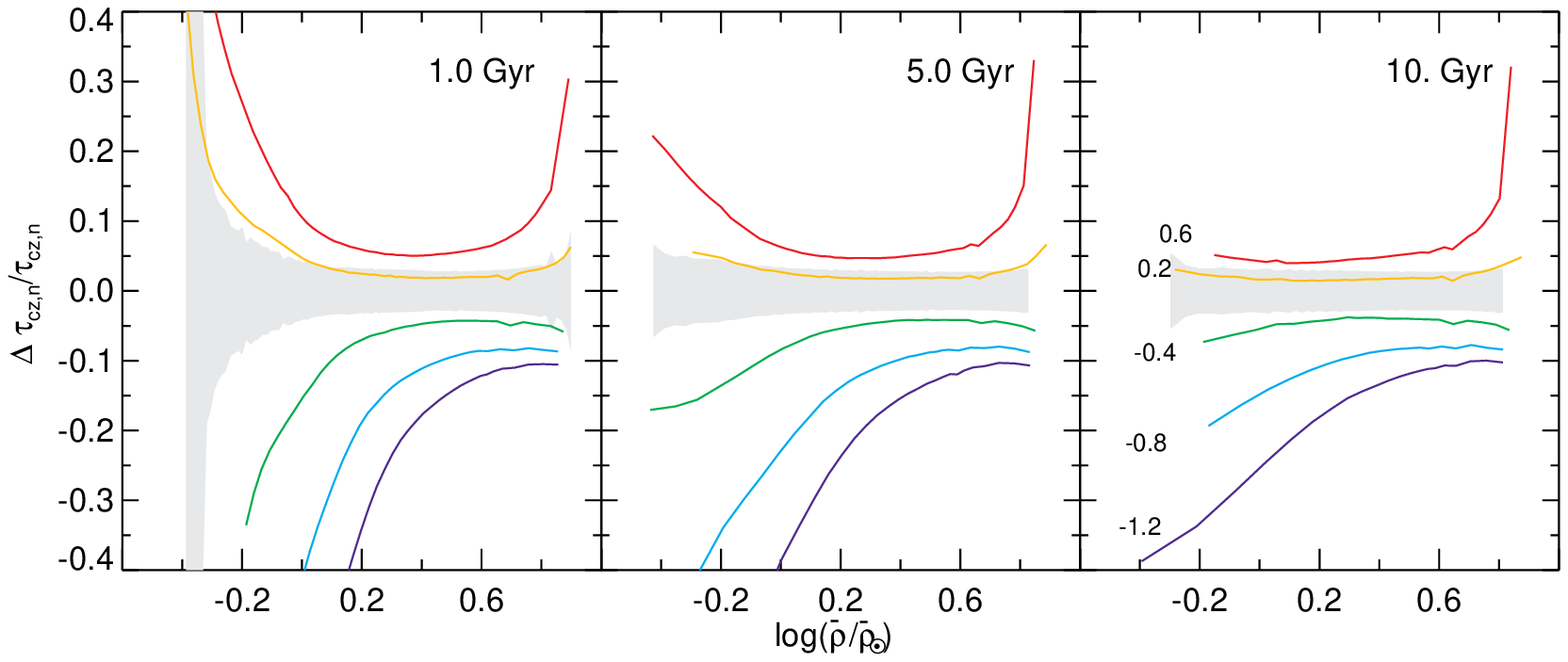}}
       \caption{The fractional difference in $\tau_{cz,n}$ (the normalized acoustic depth to the base of the convection zone) between models of different compositions as a function of $\bar{\rho}$. Each solid line represents the fractional difference in $\tau_{cz,n}$ between a given [Z/X] ($-1.2$, $-0.8$, $-0.2$, 0.2, 0.6, marked for reference in the right panel) and $\tau_{cz,n}$ for a solar composition model.  The gray shaded region represents observational and theoretical errors on $\tau_{cz,n}$, both described in detail in Sections \ref{subsec:theerror} }
	\label{fig:standard_difference_rho}
\end{figure}

\begin{figure}
	\centerline{\includegraphics[scale = 0.9]{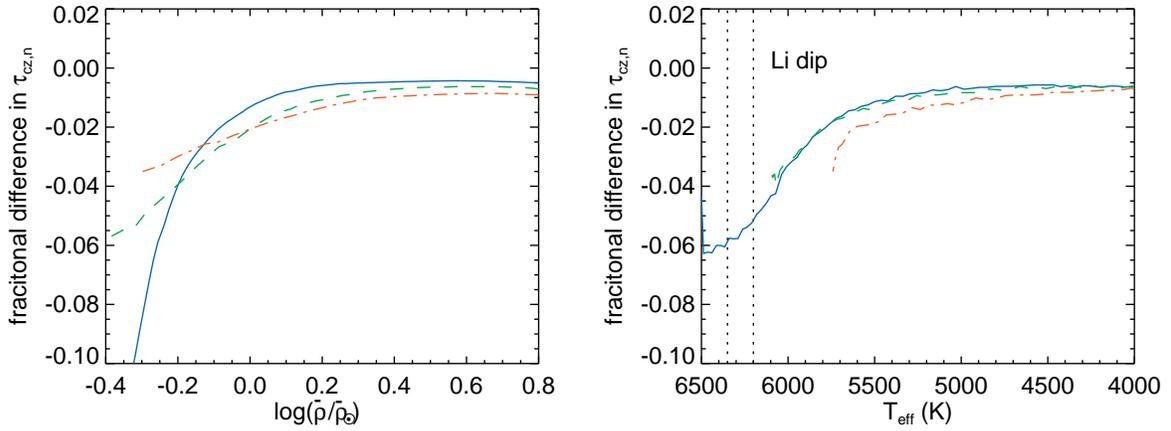}}
       \caption{Fractional difference in $\tau_{cz,n}$ between a set of models with standard physics, and a set where diffusion has been eliminated entirely to mimic efficient mixing. The solid (blue) line is for models at 1 Gyr, dashed (green) at 5 Gyr, and dot-dashed (red) at 10.0 Gyr. Note that over the temperature range in which the Li dip is observed, the difference is of order 6\%, such that over a very narrow temperature range we expect to see and aburpt change in the location of the the base of the CZ. For the 5.0 and 10.0 Gyr curves, all stars in the Li dip temperature range have already evolved off of the MS. See Section \ref{subsubsec:lidip} for further discussion}
	\label{fig:mixing_error}
\end{figure}

\begin{deluxetable}{llllll}
\tabletypesize{\scriptsize}
\tablecolumns{6}
\tablewidth{0pt}
\tablecaption{Theoretical Errorbar Model Grids}
\tablehead{
                        \colhead{Grid name}     &
                        \colhead{$\alpha$}      &
		        \colhead{$X$}		&
			\colhead{$Y$}		&
			\colhead{$Z$}		&
			\colhead{Description}   }
\startdata
standard & 1.93271 & 0.710040  & 0.018338 & 0.271622 & standard physics, Section \ref{sec:standard_physics}  \\
mixing & 1.80975 & 0.719426 & 0.016468 & 0.264107 & no diffusion \\
eos (scv) & 1.87640 & 0.709470 & 0.018455 & 0.272075 & SCV/Yale EOS \\
eos (yale) & 1.90600 & 0.716180 & 0.018665 & 0.265155 & Yale EOS only \\
overshoot & 1.91560 & 0.707060 & 0.018299 & 0.274641 & Convective core overshoot\\
mixture & 1.90600 & 0.720880 & 0.014760 &  0.264360 & \citet{asplund2009} solar mixture \\
nuclear & 1.92500 &  0.709110 & 0.018257 & 0.272633  & nuclear cross sections S11, S33, S34, S114 +4$\sigma$ \\
opacity & 1.92250 & 0.709415 & 0.018338 &  0.272247 & OPAL/Alexander opacities  \\
alpha & many & 0.710040 & 0.018338 & 0.271622 & $\alpha$ values, chosen to produce $\Delta R = +0.01$ \\
boundary & 1.82100 & 0.710030 & 0.018334 & 0.271636 & Grey atmosphere boundary condition\\
diffusion & 1.95000 & 0.708850 & 0.018597 & 0.272553 & diffusion coefficients altered by 15\%

\enddata

\label{tbl:errorbar}
\end{deluxetable}

\begin{deluxetable}{lllllllllllll}
\tabletypesize{\scriptsize}
\tablecolumns{13}
\tablecaption{Model Grid }
\tablehead{
                        \colhead{Mass \tablenotemark{a}}     &
			\colhead{$Y_i$}                 &  
                       \colhead{$[Z/X]_i$}             &
		        \colhead{$[Z/X]$}	        &
			\colhead{Age}	        &
			\colhead{$\log{(\frac{L}{L_{\odot}})}$} &	
			\colhead{$\log{(\frac{R}{R_{\odot}})}$} &
			\colhead{$T_{eff}$}             &
			\colhead{$\log{(\frac{\bar{\rho}}{\bar{\rho}_{\odot}})}$}          &
			\colhead{$R_{cz}/R$}            &
			\colhead{$\tau_{cz}$}           &
			\colhead{$\tau_{\star}$}        &
			\colhead{$\tau_{cz,n}$}          \\

                       \colhead{($M_{\odot}$)}     &
			\colhead{}                 &  
                       \colhead{(initial)}             &
		        \colhead{(surf)}	        &
			\colhead{(Gyr)}	        &
			\colhead{} &	
			\colhead{} &
			\colhead{(K)}             &
			\colhead{}          &
			\colhead{}            &
			\colhead{(s)}           &
			\colhead{(s)}        &
			\colhead{}   }

\startdata
0.400  &  0.240  &  -1.2  &  -1.20e+00  &  0.500  &  -1.611  &  -0.4313  &  3754  &  0.8961  &  0.6090  &  839  &  1237  &  0.6791  \\ 
0.400  &  0.240  &  -1.0  &  -1.00e+00  &  0.500  &  -1.630  &  -0.4306  &  3711  &  0.8940  &  0.6032  &  847  &  1240  &  0.6834  \\ 
0.400  &  0.240  &  -0.80  &  -8.01e-01  &  0.500  &  -1.652  &  -0.4308  &  3664  &  0.8946  &  0.5926  &  854  &  1238  &  0.6908  \\ 
0.400  &  0.240  &  -0.60  &  -6.01e-01  &  0.500  &  -1.676  &  -0.4328  &  3623  &  0.9005  &  0.5736  &  865  &  1229  &  0.7046  \\ 
0.400  &  0.240  &  -0.40  &  -4.01e-01  &  0.500  &  -1.697  &  -0.4361  &  3592  &  0.9105  &  0.5444  &  880  &  1214  &  0.7254  \\

\enddata
\label{tbl:model_grid}
\tablenotetext{a}{Table \ref{tbl:model_grid} is available in its entirety at www.astronomy.ohio-state.edu/CZdepth/model\_grid.txt. A portion is shown here for guidance regarding its form and content. The columns are as follows: 1) mass in solar masses, 2) initial helium abundance, 3) initial model [Z/X] referenced to the initial Z/X of a calibrated solar model, 3) the surface [Z/X] at a given age, 4) age, 5) log luminosity in solar luminosities, 6) log radius in solar radii, 7) effective temperature, 8) log mean density given by $\log{(M R^{-3}/M_{\odot} R_{\odot}^{-3})}$, 9) fractional radius of the CZ, 10) acoustic depth to the CZ, 11) acoustic crossing time, 12) normalized acoustic depth.}

\end{deluxetable}

\end{document}